\documentclass[dvips,12pt,a4paper]{article}
\usepackage{a4wide}
\usepackage{epsfig}
\usepackage{amsmath}
\usepackage{latexsym}

  \newlength{\absize}
\newcommand{\half}{{\textstyle\frac{1}{2}}}

\newcommand{\fourth}{{\textstyle\frac{1}{4}}}

\renewcommand{\Re}{\mbox{Re}}

\renewcommand{\theequation}{\thesection.\arabic{equation}}

\allowdisplaybreaks 

\catcode`@=11
\def\citer{\@ifnextchar [{\@tempswatrue\@citexr}{\@tempswafalse\@citexr[]}}
 
\def\@citexr[#1]#2{\if@filesw\immediate\write\@auxout{\string\citation{#2}}\fi
  \def\@citea{}\@cite{\@for\@citeb:=#2\do
    {\@citea\def\@citea{--\penalty\@m}\@ifundefined
       {b@\@citeb}{{\bf ?}\@warning
       {Citation `\@citeb' on page \thepage \space undefined}}%
\hbox{\csname b@\@citeb\endcsname}}}{#1}}
\catcode`@=12

\begin{document}
  \thispagestyle{empty}
  \pagestyle{empty}
  \renewcommand{\thefootnote}{\fnsymbol{footnote}}
\newpage\normalsize
    \pagestyle{plain}
    \setlength{\baselineskip}{4ex}\par
    \setcounter{footnote}{0}
    \renewcommand{\thefootnote}{\arabic{footnote}}
\newcommand{\preprint}[1]{%
  \begin{flushright}
    \setlength{\baselineskip}{3ex} #1
  \end{flushright}}
\renewcommand{\title}[1]{%
  \begin{center}
    \LARGE #1
  \end{center}\par}
\renewcommand{\author}[1]{%
  \vspace{2ex}
  {\Large
   \begin{center}
     \setlength{\baselineskip}{3ex} #1 \par
   \end{center}}}
\renewcommand{\thanks}[1]{\footnote{#1}}
\renewcommand{\abstract}[1]{%
  \vspace{2ex}
  \normalsize
  \begin{center}
    \centerline{\bf Abstract}\par
    \vspace{2ex}
    \parbox{\absize}{#1\setlength{\baselineskip}{2.5ex}\par}
  \end{center}}

\hyphenation{phenomeno-logy}
\renewcommand{\thefootnote}{\fnsymbol{footnote}}
\begin{flushright}
{\setlength{\baselineskip}{2ex}\par
{January 2004}           \\
} 
\end{flushright}
\vspace*{4mm}
\vfill
\title{Center--Edge Asymmetry at Hadron Colliders}
\vfill
\author{
E.W.\ Dvergsnes$^{a,}$\footnote{\tt erik.dvergsnes@fi.uib.no},
P.\ Osland$^{a,}$\footnote{\tt per.osland@fi.uib.no},
A.A.\ Pankov$^{b,c,}$\footnote{\tt pankov@gstu.gomel.by}
and N.\ Paver$^{c,}$\footnote{\tt nello.paver@ts.infn.it}
}
\begin{center}
$^{a}$Department of Physics and Technology, University of Bergen, 
N-5007 Bergen, Norway\\
$^{b}$Pavel Sukhoi Technical University, Gomel 246746, Belarus \\
$^{c}$University of Trieste and INFN-Sezione di Trieste, 34100 Trieste, Italy
\end{center}
\vfill

\abstract{We investigate the possibility of using the center--edge
asymmetry to distinguish graviton exchange from other new physics
effects at hadron colliders. Specifically, we study lepton-pair
production within the ADD and RS scenarios. At the Tevatron, the
graviton-$Z$ interference is the most important contribution to the
center--edge asymmetry, whereas at the LHC, the dominant contribution
comes from gluon fusion via graviton exchange, which has no analogue
at $e^+ e^-$ colliders.
We find that spin-2 and spin-1 exchange can be distinguished
up to an ADD cut-off scale, $M_H$, of about 5~TeV, at the 95\% CL.
In the RS scenario, spin-2 resonances can be identified in most of
the favored parameter space.}
\vspace*{20mm} 
\setcounter{footnote}{0} 
\vfill

\newpage
\setcounter{footnote}{0}
\renewcommand{\thefootnote}{\arabic{footnote}}

\section{Introduction}\label{sec:I}
\setcounter{equation}{0}
There exists a variety of proposals of what may come beyond the Standard Model
(SM), many different scenarios introduce new fundamental particles and forces
at very high mass scales.  Some of these different proposals are: composite
models of quarks and leptons \cite{'tHooft:xb,Eichten:1983hw}; exchanges of
heavy $Z'$ \cite{Barger:1997nf,Hewett:1988xc} and (scalar and vector)
leptoquarks \cite{Buchmuller:1986zs}; $R$-parity breaking sneutrino exchange
\cite{Kalinowski:1997bc,Rizzo:1998vf}; 
anomalous gauge boson couplings \cite{Gounaris:1997ft};
Kaluza--Klein (KK) graviton exchange, exchange of gauge boson KK towers or
string excitations, {\it etc.}  \citer{Hewett:1999sn,Dienes:1998vh}.
There is a hope that new physics (NP) effects will be observed either
directly, as in the case of new particle production, e.g., $Z'$ and $W'$
vector bosons, SUSY or Kaluza-Klein (KK) resonances, or indirectly through
deviations, from the SM predictions, of observables such as cross sections and
asymmetries.

Over the last years, intensive studies have been carried out, of how different
scenarios involving extra dimensions
\citer{Arkani-Hamed:1998rs,Antoniadis:1990ew} would manifest themselves at
high energy colliders such as the Tevatron, the Large Hadron Collider (LHC)
and an $e^+e^-$ Linear Collider (LC) \citer{Hewett:1999sn,Dienes:1998vh}.  We
shall consider the possibility of distinguishing such effects of extra
dimensions from other NP scenarios at hadron colliders, focusing on two
specific models involving extra dimensions, namely the
Arkani-Hamed--Dimopoulos--Dvali (ADD) \cite{Arkani-Hamed:1998rs} and
Randall--Sundrum (RS) \cite{Randall:1999ee} scenarios with emphasis on the
lepton pair production process.  These models lead to very different
phenomenologies and collider signatures.

The large extra dimension scenario (ADD) predicts the emission and exchange of
KK towers of gravitons. The effect of the graviton towers can be described
through a set of dimension-8 operators characterized by a
large cut-off scale, $M_H$
\cite{Hewett:1999sn}.  The distortion of the differential Drell-Yan cross
section at large values of the dilepton invariant mass through these
dimension-8 operators can probe such high mass scales in a manner similar to
searches for contact interactions in composite models. The shape of the
invariant mass distribution will tell us that the underlying physics arises
from dimension-8 operators, while the angular distribution of the leptons at
large dilepton invariant masses would have the shape expected from the
exchange of a spin-2 object, confirming the gravitational origin of the
effect.

The phenomenology of the RS model with warped extra dimensions is very
different from the ADD model in two aspects: (i) the spectrum of the graviton
KK states are discrete and unevenly spaced while it is uniform, evenly spaced,
and effectively a continuous spectrum in the ADD model, and (ii) each
resonance in the RS model has a coupling strength of 1/TeV while in the ADD
model only the collective strength of all graviton KK states gives a coupling
strength 1/TeV. The RS model predicts TeV-scale graviton resonances which
might be produced in many channels, including the dilepton channel. The spin-2
nature of the graviton resonance can be determined from the distinct shape of
the angular distribution of the final state leptons in Drell-Yan production at
the Tevatron and LHC.

Many different NP scenarios may lead to the same or very similar experimental
signatures. Therefore, searching for effects of extra dimensions can be
jeopardized by the misidentification of their signal with other possible
sources of new phenomena. Thus, it is important to study how to differentiate
the corresponding signals.

One can develop techniques which will help dividing models into distinct
subclasses.  In this paper we shall discuss a technique that makes use of the
specific modifications in angular distributions induced by spin-2 exchanges.
This method is based on the center-edge asymmetry $A_\text{CE}$
\cite{Osland:2003fn,Pankov:1997da}, an integrated observable which offers a
way to uniquely identify KK graviton exchange (or any other spin-2 exchange).
In a situation with limited statistics, this may represent an advantage over a
fit to the angular distribution.

In Sec.~\ref{sec:no-cuts} we define a hadron collider version of the
center--edge asymmetry, and give the cross sections relevant for lepton
production.  Thereafter, in Sec.~\ref{sec:cuts}, we study the effects of
introducing angular cuts related to the detector geometries.  In
Sec.~\ref{sec:identify} we then
discuss how the center--edge asymmetry can be used to identify spin-2 exchange
at both the LHC and the Tevatron. Finally, in Sec.~\ref{sec:summary} we
summarize our results.

\section{The center--edge asymmetry $A_\text{CE}$}\label{sec:no-cuts}
\setcounter{equation}{0}

At hadron colliders, lepton pairs can in the SM be produced at tree-level via
the following sub-process
\begin{equation}
q \bar q \to \gamma,Z \to l^+ l^-,
\end{equation}
where we shall use $l=e,\mu$.
If gravity can propagate in extra dimensions, the possibility of KK graviton 
exchange opens up two tree-level channels at hadron colliders in addition to 
the SM channels, namely
\begin{align}
q \bar q \to G \to l^+ l^-, \nonumber \\
gg \to G \to l^+ l^-,
\end{align}
where $G$ represents the gravitons of the KK tower. At the LHC, the 
gluon-fusion channel can give an important contribution, since it has a 
different angular distribution arising from the difference between 
the gluon-graviton and quark-graviton couplings, combined with the high gluon 
luminosities.

Consider a lepton pair of invariant mass $M$ at rapidity $y$ (of the parton
c.m.\ frame) and with $z=\cos\theta_\text{cm}$, where $\theta_\text{cm}$ is
the angle, in the c.m.\ frame of the two leptons, between the lepton ($l^-$)
and the proton $P_1$.  The inclusive differential cross section for producing
such a pair, can at the LHC proton-proton collider be expressed as
\begin{align}
\label{Eq:dsigma-dMdydz}
&\begin{aligned}
\frac{d\sigma_{q \bar q}}{dM\,dy\,dz}
= K\frac{2 M}{s} \sum_q \biggl\{&[f_{q|P_1}(\xi_1,M)f_{\bar q|P_2}(\xi_2,M)
+ f_{\bar q|P_1}(\xi_1,M)f_{q|P_2}(\xi_2,M)]
\frac{d\hat \sigma^\text{even}_{q \bar q}}{dz} \nonumber \\
+&[f_{q|P_1}(\xi_1,M)f_{\bar q|P_2}(\xi_2,M)
- f_{\bar q|P_1}(\xi_1,M)f_{q|P_2}(\xi_2,M)]
\frac{d\hat \sigma^\text{odd}_{q \bar q}}{dz}\biggr\},  \nonumber \\
\end{aligned} \\
&\begin{aligned}
\frac{d\sigma_{gg}}{dM\,dy\,dz}
= K\frac{2 M}{s} \, f_{g|P_1}(\xi_1,M)f_{g|P_2}(\xi_2,M)
\frac{d\hat\sigma_{gg}}{dz}.
\end{aligned}
\end{align}
Here, $d\hat\sigma_{q \bar q}^\text{even}/dz$ and $d\hat\sigma_{q
\bar q}^\text{odd}/dz$ are the even and odd parts (under
$z\leftrightarrow -z$) of the partonic differential cross
section $d\hat\sigma_{q \bar q}/dz$, and the minus sign in
the odd term allows us to interpret the angle in the parton cross section
as being relative to the quark momentum (rather than $P_1$).
Furthermore, $K$ is a factor accounting for
higher order QCD corrections (we take $K=1.3$, which is a typical value),
$f_{j|P_i}(\xi_i,M)$ are parton distribution functions in the proton
$P_i$, and the $\xi_i$ are fractional parton momenta
\begin{equation}
\xi_1=\frac{M}{\sqrt{s}}e^y, \qquad 
\xi_2=\frac{M}{\sqrt{s}}e^{-y}.
\end{equation}
We also made use of the relation $d\xi_1\,d\xi_2=dM(2M/s)dy$
and have $M^2=\xi_1\xi_2 s$, with $s$ the $pp$ c.m.\ energy squared.

At the Tevatron, taking into account that one beam consists
of antiprotons, the following substitution must be made
in (\ref{Eq:dsigma-dMdydz}):
\begin{equation} \label{pbar-p}
P_1\to P, \quad P_2 \to \bar P. 
\end{equation}

The center--edge and total cross sections can at the parton level be defined
like for initial-state electrons and positrons
\cite{Osland:2003fn,Pankov:1997da}:
\begin{equation}
\label{Eq:sigma-hat-ce}
\hat \sigma_\text{CE}
\equiv \left[\int_{-z^*}^{z^*} 
- \left(\int_{-1}^{-z^*}
+\int_{z^*}^{1}\right)\right]
\frac{d\hat \sigma}{dz}\, dz, \quad
\hat \sigma
\equiv \int_{-1}^{1} 
\frac{d\hat \sigma}{dz}\, dz.
\end{equation}
These will play a central role in the center--edge asymmetry
at the hadron level.  At this point, $0<z^*<1$
is just an arbitrary parameter which defines the border between the ``center''
and the ``edge'' regions.

At hadron colliders, the center--edge asymmetry can 
for a given dilepton invariant mass $M$ be defined as 
\begin{equation}
\label{Eq:ace}
A_\text{CE}(M)=\frac{d\sigma_\text{CE}/dM}
                 {d\sigma/dM},
\end{equation}
where we obtain $d\sigma_\text{CE}/dM$ and $d\sigma/dM$ from 
(\ref{Eq:dsigma-dMdydz}) by integrating over $z$ according
to Eq.~(\ref{Eq:sigma-hat-ce}) and over rapidity between
$-Y$ and $Y$, 
with $Y=\log(\sqrt{s}/M)$. Furthermore [see Eq.~(\ref{Eq:dsigma-dMdydz})],
\begin{equation}
\frac{d\sigma}{dM}
=\frac{d\sigma_{q \bar q}}{dM}
+\frac{d\sigma_{gg}}{dM}.
\end{equation}

We note that terms in the parton cross sections that are odd in
$z$ do not contribute to $A_\text{CE}$; and that
\begin{equation}
\label{Eq:CE-limits}
\frac{d\sigma_\text{CE}}{dM}\bigg|_{z^*=0}
=-\frac{d\sigma}{dM}, \qquad
\frac{d\sigma_\text{CE}}{dM}\bigg|_{z^*=1}
=\frac{d\sigma}{dM}.
\end{equation}

Conversely, the odd terms of the partonic cross section determine the 
familiar forward-backward asymmetry $A_\text{FB}$:
\begin{equation}
\left(\frac{d\sigma}{dM}\right)A_\text{FB}(M)=
\left(\int_{0}^{Y} \mp \int_{-Y}^0\right)\,dy\,
\left(\int_0^1-\int_{-1}^0\right)\,dz\,
{\frac{d\sigma}{dM\,dy\,dz}},
\label{Eq:AFB}
\end{equation}
where the two signs in the integration over $y$ refer to the proton-proton and
proton-antiproton cases, respectively. Note that on the right-hand side of
Eq.~(\ref{Eq:AFB}) the gluon-gluon channel will not contribute, as it is
symmetric in $z$.

In order to develop some intuition for the different contributions to the
cross section, and to the asymmetry $A_\text{CE}(M)$, we shall first study the
ideal case where no angular cuts have been imposed. The modifications caused
by quasi-realistic cuts will be discussed in Sec.~\ref{sec:cuts}.

\subsection{$A_\text{CE}$ in the ADD Scenario}\label{subsec:add}
Let us now consider the ADD scenario \cite{Arkani-Hamed:1998rs}, where gravity
is allowed to propagate in two or more compactified, but still large, extra
dimensions. This gives rise to a tower of (massive) KK gravitons with tiny
mass splittings. In the Hewett approach \cite{Hewett:1999sn,Hewett:2002hv},
the summation over KK states (of mass $m_{\vec n}$) is performed by the
following substitution:
\begin{equation}
\label{Eq:Hewett}
\sum_{\vec n =1}^\infty \frac{G_\text{N}}{M^2 - m_{\vec n}^2} 
\to \frac{-\lambda}{\pi\,M_H^4},
\end{equation}
where $\lambda$ is a sign factor, and $G_\text{N}$ is Newton's constant.  This
approach takes into account the fact that the ultraviolet behavior of the
scenario is unknown (for a recent discussion, see \cite{Giudice:2003tu}). In
particular, there is no dependence on the number of extra dimensions.

We then have the following parton differential cross sections
\cite{Gupta:1999iy}, where double superscripts refer to interference between
the respective amplitudes (with $z$ the cosine of the quark-lepton angle in
the dilepton c.m. frame, and averaged over quark and gluon colors):
\begin{alignat}{2}
\label{Eq:sigma}
\frac{d\hat \sigma_{gg}^G}{dz}
&=\frac{\lambda^2 M^6}{64 \pi M_H^8}(1-z^{4}),& \qquad
\frac{d\hat \sigma_{q\bar q}^G}{dz}
&=\frac{\lambda^2 M^6}{96 \pi M_H^8}
(1-3z^{2}+4z^{4}), 
\nonumber \\
\frac{d\hat \sigma_{q\bar q}^{G \gamma}}{dz}
&=-\frac{\lambda \alpha Q_q Q_e M^2}{6 M_H^4}z^3,& \qquad
\frac{d\hat \sigma_{q\bar q}^{G Z}}{dz}
&=\frac{\lambda \alpha M^2}{12 M_H^4}
[a_q a_e(1-3z^{2}) - 2 v_q v_e z^3]\Re\, \chi, 
\nonumber \\
\frac{d\hat \sigma_{q\bar q}^\text{SM}}{dz}
&=\frac{\pi \alpha^2}{6 M^2}[S_q\,
(1+z^{2})+2 A_q\, z].
\end{alignat}
Here, fermion masses are neglected, and we define
\begin{align}
S_q&\equiv Q_q^2 Q_e^2 + 2 Q_q Q_e v_q v_e\, \Re\, \chi 
+(v_q^2+a_q^2)(v_e^2+a_e^2)\,|\chi|^2, \nonumber \\
A_q&\equiv 2 Q_q Q_e a_q a_e \Re\, \chi + 4 v_q a_q v_e a_e |\chi|^2.
\end{align}
We use a convention where $a_f=T_f$,
$v_f=T_f - 2Q_f \sin^2\theta_W$ and the $Z$ propagator is represented by
\begin{equation}
\chi=\frac{1}{\sin^2(2 \theta_W)}\, 
\frac{M^2}{M^2 - m_Z^2 + i m_Z \Gamma_Z}.
\end{equation}

From Eqs.~(\ref{Eq:sigma-hat-ce}) and (\ref{Eq:sigma}), we obtain the
following parton level center--edge cross sections
\begin{alignat}{2}
\label{Eq:sigma_ce}
\hat \sigma_{gg,\text{CE}}^G
&=\frac{\lambda^2 M^6}{40 \pi M_H^8}
[\half z^*(5-z^{*\,4})-1],& \qquad
\hat \sigma_{q \bar q,\text{CE}}^G
&=\frac{\lambda^2 M^6}{60 \pi M_H^8}[2z^{*\,5}
+{\textstyle\frac{5}{2}}z^*(1-z^{*\,2})-1], \nonumber \\
\hat \sigma_{q \bar q,\text{CE}}^{G \gamma}&=0,& \qquad
\hat \sigma_{q \bar q,\text{CE}}^{G Z}
&=\frac{\lambda \alpha a_q a_e M^2}{3 M_H^4} \, 
\Re\, \chi [z^*(1-z^{*\,2})], \nonumber \\
\hat \sigma_{q \bar q,\text{CE}}^\text{SM}
&=\frac{4\pi \alpha^2}{9 M^2}\,S_q\,
[\half z^*(z^{*\,2}+3)-1].
\end{alignat}
Note that, in contrast to the integrated cross section, where both $\hat
\sigma_{q \bar q}^{G\gamma}$ and $\hat \sigma_{q \bar q}^{GZ}$ vanish, the
interference between graviton and $Z$ exchange may play an important role in
enhancing the sensitivity to the graviton exchange. Actually, one should
observe from Eqs.~(\ref{Eq:sigma}) and (\ref{Eq:sigma-hat-ce}) that such
interference is suppressed, for $M/M_H<1$, by a factor $(M/M_H)^4$ relative to
the SM, while all ``pure'' graviton exchange contributions are suppressed by
the more severe factor $(M/M_H)^8$. Therefore, in particular, the gluon-gluon
channel is expected to give a significant contribution only at the LHC
collider thanks to the large gluon-gluon luminosities available there.

For the SM contribution to the center--edge asymmetry, we see that the
convolution integrals, depending on the parton distribution functions, cancel,
and the result is
\begin{equation}
\label{Eq:ACEspin1}
A_\text{CE}^\text{SM}
=\half\, z^*(z^{*2}+3)-1,
\end{equation} 
which is independent of $M$ and identical to the result for $e^+e^-$ colliders
\cite{Osland:2003fn}.  Hence, in the case of no cuts, there is a unique value,
$z_0^*$, of $z^*$ for which $A_\text{CE}^\text{SM}$ vanishes
\cite{Datta:2002tk}:
\begin{equation} 
\label{Eq:z0*}
z_0^* = (\sqrt{2} + 1)^{1/3} - (\sqrt{2} - 1)^{1/3} \simeq 0.596,
\end{equation} 
corresponding to $\theta_\text{cm} = 53.4^\circ$.

The structure of the differential SM cross section of Eq.~(\ref{Eq:sigma}) is
particularly interesting in that it is equally valid for a wide variety of NP
models: composite-like contact interactions, $Z'$ models, TeV-scale gauge
bosons, {\it etc}.
Conventional four-fermion contact-interaction effects of the vector--vector
kind would yield the same center--edge asymmetry as the SM.  If however KK
graviton exchange is possible, the tensor couplings would yield a different
angular distribution, hence a different dependence of $A_\text{CE}$ on
$z^*$. In particular, the center--edge asymmetry would not vanish
for the same choice of $z^* = z_0^*$ and, moreover, would show a 
non-trivial dependence on $M$. Thus, a value for
$A_\text{CE}$ different from $A_\text{CE}^\text{SM}$ would 
indicate non-vector exchange NP.

The other important difference from the spin-1 exchange originating from
$q\bar q$ annihilation is that the graviton also couples to gluons, and
therefore, it has the additional $gg$ initial state available, see
Eq.~(\ref{Eq:sigma}). As a result of including graviton exchange, the
center--edge asymmetry is no longer the simple function of $z^*$ in
Eq.~(\ref{Eq:ACEspin1}).
\subsubsection{Parton-level asymmetry}

We start the quantitative discussion of the $A_\text{CE}$ asymmetry by
considering a simple, limiting case which illustrates, at the parton level,
the $M$-behavior of the $gg$ vs.\ $q\bar q$ subprocesses and the
corresponding interference with the SM.  In Fig.~\ref{Fig:aceadd-shat-parton}
we show the `parton-level' quantity
\begin{equation}
\widehat A_\text{CE} \equiv \frac{\hat \sigma_\text{CE}}{\hat \sigma},
\end{equation}
where
\begin{equation}
\hat \sigma_\text{CE} 
= \hat \sigma_{u\bar u,\text{CE}}^G
+ \hat \sigma_{gg,\text{CE}}^G
+ \hat \sigma_{u\bar u,\text{CE}}^{GZ},
\end{equation}
and
\begin{equation}
\hat \sigma = \hat \sigma_{u\bar u}^G
+ \hat \sigma_{gg}^G + \hat \sigma_{u\bar u}^\text{SM}.
\end{equation}
Thus, in this example, we take the `protons' to consist of only $u$ quarks,
$u$ antiquarks and gluons. In a collision among such `protons',
we also require the probability of finding a $u\bar u$ pair to be equal to the
probability of finding a $gg$ pair (the respective convolution integrals are
set equal to unity), independent of the invariant mass $M$. In this limit, the
mass dependence arises solely from the parton-level cross sections.
Furthermore, we consider $z^*=z_0^*\simeq0.596$, such that $\hat \sigma_{u\bar
u,\text{CE}}^\text{SM}=0$.  We note that in the limit $M\gg M_Z$, the
contributions to $\widehat A_\text{CE}$ given in
Fig.~\ref{Fig:aceadd-shat-parton} are of the form $f(M/M_H)$.

\begin{figure}[htb]
\refstepcounter{figure}
\label{Fig:aceadd-shat-parton}
\addtocounter{figure}{-1}
\begin{center}
\setlength{\unitlength}{1cm}
\begin{picture}(7.7,7.7)
\put(0.0,0.0)
{\mbox{\epsfysize=8.0cm\epsffile{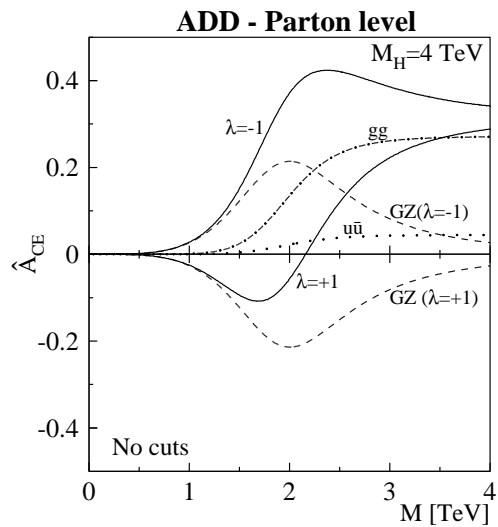}}}
\end{picture}
\caption{Different contributions to $\widehat A_\text{CE}$ as a function of
$M$ in the ADD scenario, for $M_H=4\text{ TeV}$.  A simplified parton-level
situation is considered, see text.  Solid curves: total (parton-level)
center--edge asymmetry ($\lambda=\pm 1$), dash-dotted: $gg$ contribution,
dotted: $u \bar u$ with graviton exchange, dashed: $u \bar u$ interference
between graviton and $Z$ (labeled $GZ$).}
\end{center}
\end{figure}

Since $\hat \sigma_{qq,\text{CE}}^{GZ} \propto a_e a_q$, a cancellation would
occur if both $d$- and $u$-quarks ($a_d=-a_u$) were considered with equal
weight. This cancellation is only partial when differences in parton
distributions are accounted for.

For $z^*=z_0^*$ as suggested by
Eq.~(\ref{Eq:z0*}), and considering the limits of pure
glue--glue and pure $q\bar q$, we find:
\begin{align} \label{Eq:parton-CE-limits}
\widehat A_{gg,\text{CE}}^G
=&\frac{\hat\sigma_{gg,\text{CE}}^{G}}{\hat\sigma_{gg}^{G}}
=\half\,z_0^*(5-z_0^{*\,4})-1 \simeq 0.453, \nonumber \\
\widehat A_{q\bar q,\text{CE}}^{G}
=&\frac{\hat\sigma_{q\bar q,\text{CE}}^{G}}{\hat\sigma_{q\bar q}^{G}}
=2z_0^{*\,5}
+\textstyle{\frac{5}{2}}\,z_0^*(1-z_0^{*\,2})-1\simeq 0.111,
\end{align}
which shows that gluon-fusion events are more ``centered'' than
quark-antiquark annihilation events with graviton exchange.  Furthermore,
since both quantities in (\ref{Eq:parton-CE-limits}) are positive, pure
graviton events are in general more centered than SM events.

From the cross section formulas (\ref{Eq:sigma_ce}) [see also
(\ref{Eq:CE-limits})] we see that $\hat \sigma_{gg}^G = (3/2)\hat
\sigma_{u\bar u}^G$.  At very large $M$, the SM result, and therefore also the
interference, will be negligible. In this case, $\hat \sigma_{gg}^G$ and
$\hat\sigma_{u\bar u}^G$ will contribute 60\% and 40\% to the cross section.
Therefore, for large $M$, using the values of Eq.~(\ref{Eq:parton-CE-limits}),
we find that $\widehat A_\text{CE} \to 0.6\times0.453 +0.4\times0.111 =
0.314$, in agreement with Fig.~\ref{Fig:aceadd-shat-parton}.  This limit is
also applicable at the peak of an RS-graviton resonance (where the
interference vanishes). However, as we shall see below, when parton
distributions are included, the simple relation 3:2 between the cross sections
for gluon fusion and quark-antiquark annihilation with graviton exchange is no
longer valid.
\subsubsection{Including parton distributions}
When parton distributions are included, the picture changes,
since the relative probabilities of finding a $gg$ pair and
a $q\bar q$ pair of invariant mass $M$ will depend on $M$,
as well as on the collision energy, and whether one considers the
Tevatron ($p\bar p$) or the LHC ($pp$).
In Fig.~\ref{Fig:aceadd-shat-cont} we show $A_\text{CE}$ (for
$z_0^*\simeq0.596$) in the ADD model as a function of invariant dilepton
mass, $M$. In the left panel we consider $p \bar p$
collisions, where we have chosen the cut-off $M_H=1.4$~TeV,
$\lambda=\pm 1$ and $\sqrt{s}=1.96$~TeV (Tevatron), whereas in the
right panel $pp$ collisions are considered, with $M_H=4$~TeV,
$\lambda=\pm 1$ and $\sqrt{s}=14$~TeV (LHC). In both plots, the SM
contribution, $A_\text{CE}^\text{SM}$, to the center--edge asymmetry
vanishes.  To compute cross sections we use the CTEQ6 parton
distributions \cite{Pumplin:2002vw}.
\begin{figure}[htb]
\refstepcounter{figure}
\label{Fig:aceadd-shat-cont}
\addtocounter{figure}{-1}
\begin{center}
\setlength{\unitlength}{1cm}
\begin{picture}(16.2,7.7)
\put(0.0,0.0)
{\mbox{\epsfysize=8.0cm\epsffile{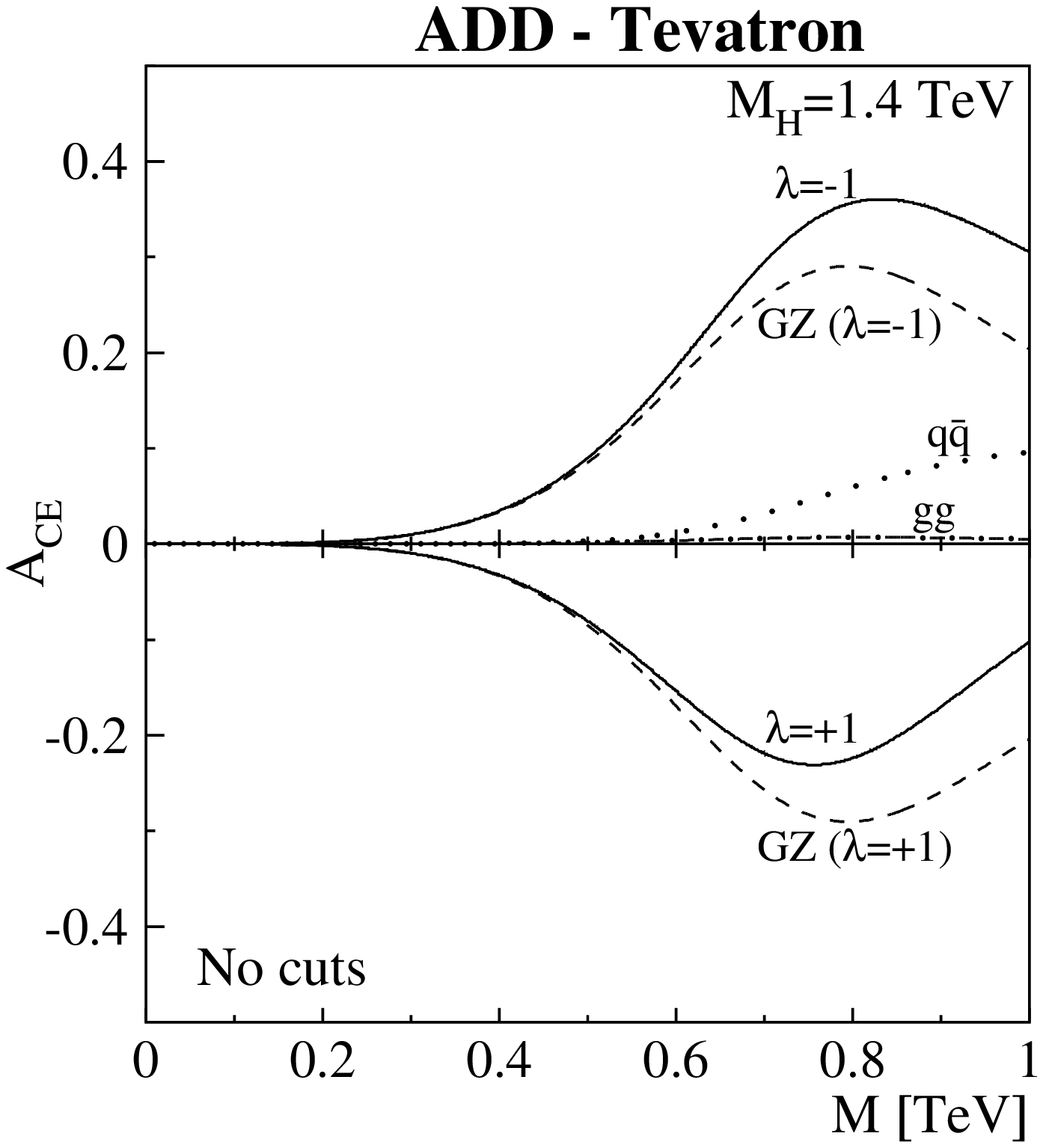}}
 \mbox{\epsfysize=8.0cm\epsffile{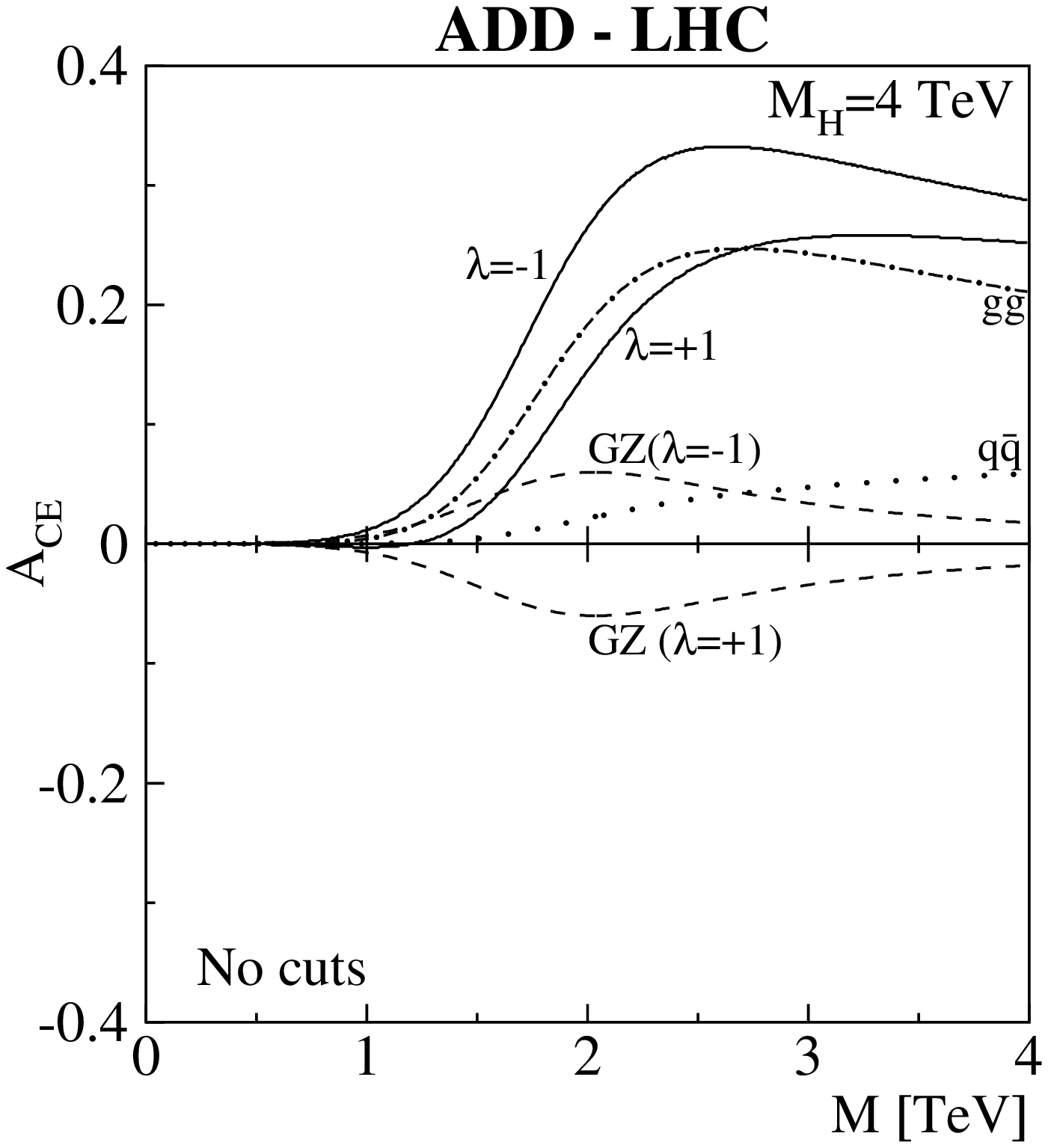}}}
\end{picture}
\caption{Different contributions to $A_\text{CE}(M)$
in the ADD scenario.
Left panel: Tevatron, $\sqrt{s}=1.96$~TeV;
Right panel: LHC, $\sqrt{s}=14$~TeV. 
Solid curves: total center--edge asymmetry ($\lambda=\pm 1$),
dash-dotted: $gg$ contribution, dotted: $q \bar q$ with graviton exchange,
dashed: $q \bar q$ interference between graviton and $Z$ (labeled $GZ$).}
\end{center}
\end{figure}

At the Tevatron, it is the graviton-$Z$ interference term which is dominant,
whereas the gluon-fusion channel is almost negligible at this energy.  We note
that the interference between the graviton and $Z$ amplitude has opposite sign
compared to that which occurs in the process $e^+e^-\to l^+l^-$
\cite{Osland:2003fn}.  This is because there is a difference in sign between
the axial vector couplings $a_u$ and $a_\ell$, $u \bar u \to l^+l^-$ being the
most important initial-state $q\bar q$ channel.  From Eq.~(\ref{Eq:sigma_ce})
it is obvious that only the graviton-$Z$ interference term is affected by the
choice of $\lambda$.

The situation is quite different at a $pp$ collider like the LHC. Here
we see that the contribution from gluon fusion (dash-dotted) actually
is the most important one. As a result, $A_\text{CE}$ becomes positive
at large $M$, independent of the sign of $\lambda$.
\subsection{$A_\text{CE}$ in the RS Scenario}\label{subsec:rs}
Another scenario involving extra dimensions, is the RS scenario
\cite{Randall:1999ee}.  Here we shall consider the simplest version of this
scenario, with only one extra dimension. The main difference from the ADD
scenario is that there will be narrow graviton resonances with masses of the
order of TeV, with couplings comparable to weak couplings.

In the RS scenario, the spacing between KK resonances can give some hints to
the underlying physics.  However, it is conceivable that the second resonance
would be outside the accessible range of the experiment, such that only the
first one would be discovered. It would then be of great interest to determine
whether it is a graviton resonance or something less exotic, like a $Z'$ with
vector couplings.

This model has two independent parameters, which we take to be $k/\bar
M_\text{Pl}$ and $m_1$, where $k$ is a constant of ${\cal O}(\bar
M_\text{Pl})$ ($k/\bar M_\text{Pl}$ is in the range $0.01$ to $0.1$
\cite{Davoudiasl:2000jd}), and $m_1$ is the mass of the first graviton
resonance.
The summation over KK states is performed without using the
substitution in Eq.~(\ref{Eq:Hewett}), but instead modifying in the left-hand side 
the graviton coupling to matter
\begin{equation}
G_\text{N} \to \frac{x_1^2}{8 \pi m_1^2}
               \left(\frac{k}{\bar M_\text{Pl}}\right)^2,
\end{equation}
while keeping the sum over propagators. Here,
$x_1=3.8317$ is the first root of the Bessel function $J_1(x_n)=0$
\cite{Randall:1999ee,Davoudiasl:2000jd}.

The deviations of $A_\text{CE}$ from the SM value (which is zero for
$z^*=z^*_0$ as defined in (\ref{Eq:z0*}), still without introducing any cuts)
are localized in the invariant mass of the lepton pair around the resonance
mass, as is illustrated in Fig.~\ref{Fig:RS-nc} for $k/\bar M_\text{Pl}=0.05$
and $m_1=500$~GeV ($m_1=2.5$~TeV) at the Tevatron (LHC). For this choice of
parameters, it is unlikely that the second resonance will be discovered. We
have used the definition of $A_\text{CE}$ given in Eq.~(\ref{Eq:ace}).

\begin{figure}[htb]
\refstepcounter{figure}
\label{Fig:RS-nc}
\addtocounter{figure}{-1}
\begin{center}
\setlength{\unitlength}{1cm}
\begin{picture}(16.2,7.7)
\put(0.0,0.0)
{\mbox{\epsfysize=8.0cm\epsffile{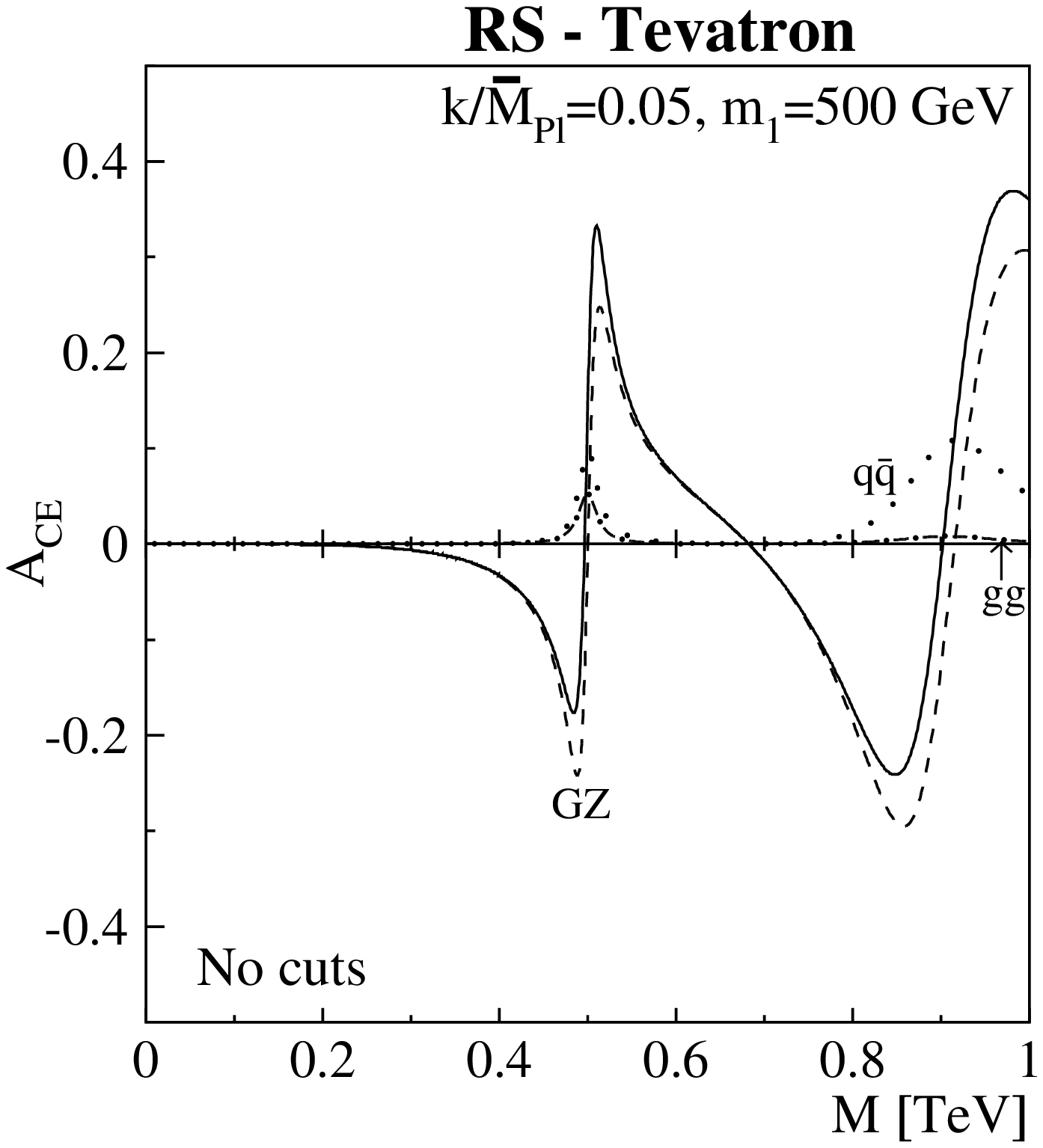}}
 \mbox{\epsfysize=8.0cm\epsffile{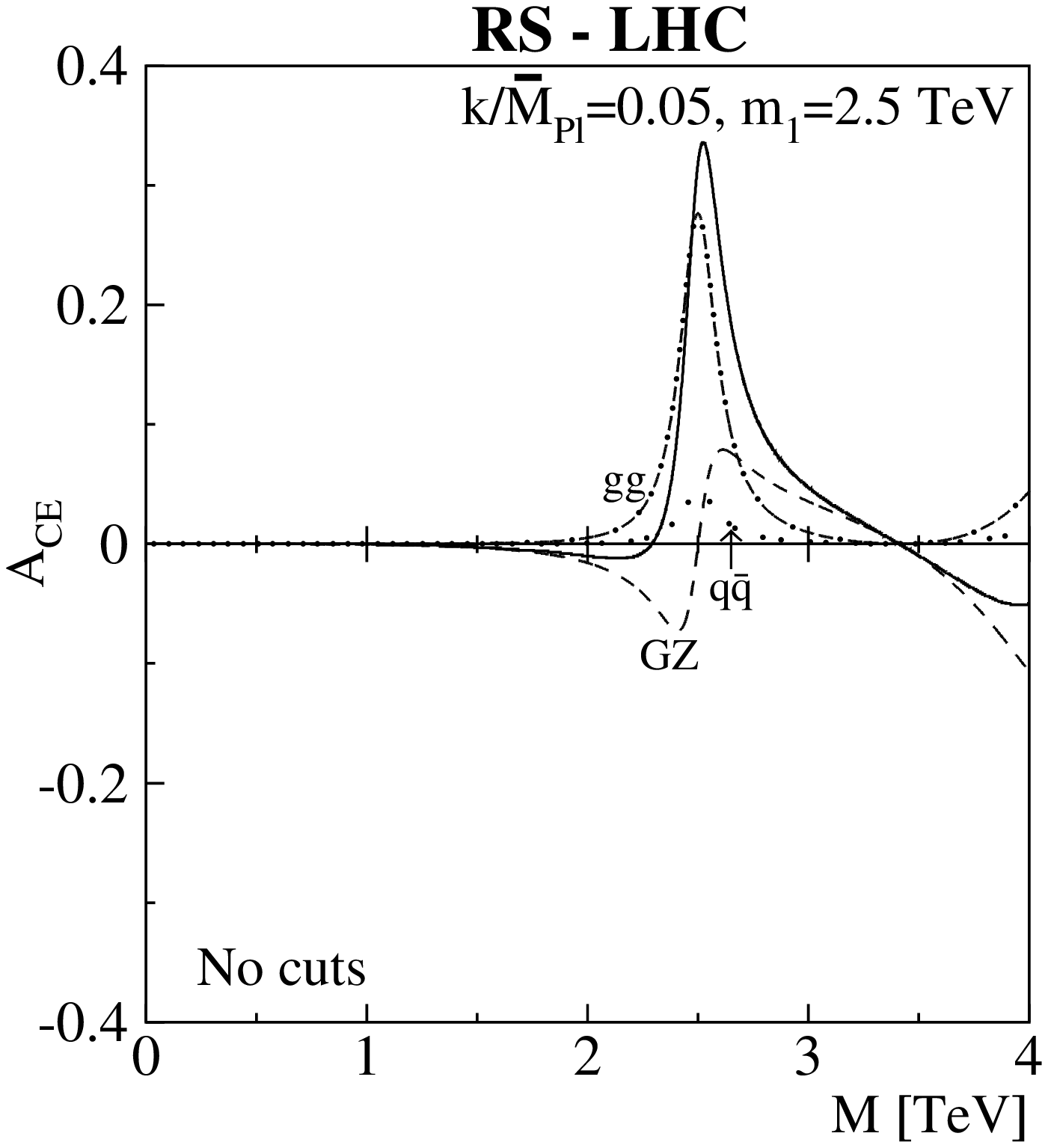}}}
\end{picture}
\caption{Different contributions to $A_\text{CE}(M)$
in the RS scenario, with $k/\bar M_\text{Pl}=0.05$.
Left panel: Tevatron, $\sqrt{s}=1.96$~TeV, $m_1=0.5$~TeV;
Right panel: LHC, $\sqrt{s}=14$~TeV, $m_1=2.5$~TeV. 
Solid curves: total center--edge asymmetry,
dash-dotted: $gg$ contribution, dotted: $q \bar q$ with graviton exchange,
dashed: $q \bar q$ interference between graviton and $Z$.}
\end{center}
\end{figure}

The graviton-$Z$ interference term vanishes, and thus changes sign, at the
resonance $M=m_1$.  At ``low'' energies, this interference gives the dominant
contribution to $A_\text{CE}$ (see left panel of Fig.~\ref{Fig:RS-nc}).  This
implies that if one integrates over some region in $M$ around the resonance,
there will be a strong cancellation.  In fact, the contribution from
graviton-SM interference almost vanishes when integrated symmetrically around
the resonance in the RS scenario, reducing the search reach for gravitons in
the RS model at the Tevatron, when using the center--edge asymmetry. This is
not the situation for the high energy attained at LHC, dominated in practice
by the $gg$ process with graviton exchange, see Fig.~\ref{Fig:RS-nc}.

\section{Introducing cuts}\label{sec:cuts}
\setcounter{equation}{0}
We next introduce angular cuts, in order to account for the fact that
detectors have a region of reduced or no efficiency close to the beam
direction.  Thus, $d\sigma/dM$ and $d\sigma_\text{CE}/dM$ of
Eq.~(\ref{Eq:ace}) are replaced by
\begin{align}
\label{Eq:sigma_ce_def_cm}
\frac{d\sigma}{dM}
&= \int_{-y_\text{max}}^{y_\text{max}}dy
\int_{-z_\text{cut}}^{z_\text{cut}}
\frac{d\sigma}{dM\,dy\,dz}dz, \nonumber \\
\frac{d\sigma_\text{CE}}{dM}
&= \int_{-y_\text{max}}^{y_\text{max}}dy
\left[\int_{-z^*}^{z^*} 
- \left(\int_{-z_\text{cut}}^{-z^*}
+\int_{z^*}^{z_\text{cut}}\right)\right]
\frac{d\sigma}{dM\,dy\,dz}dz, \quad \text{if }
z^*<z_\text{cut}, \nonumber \\
\frac{d\sigma_\text{CE}}{dM} &= \frac{d\sigma}{dM}, \quad \text{if }
z^*>z_\text{cut},
\end{align}
where the $z_\text{cut}$ is determined by detector capabilities,
and $y_\text{max}$ may be less than $Y=\log(\sqrt{s}/M)$ as a consequence
of the angular cuts.
When translated from the laboratory frame to the dilepton c.m.\ frame, the
angular cuts become boost dependent. Hence,
$z_\text{cut}=z_\text{cut}(y)$.
\subsection{LHC}\label{subsec:cut-lhc}
At the LHC, the lepton pseudorapidity cut is $|\eta| < \eta_\text{cut}=2.5$
for both leptons \cite{LHC-ref}, which corresponds to
$-z_\text{cut}^\text{lab} < z^\text{lab} < z_\text{cut}^\text{lab}$ (with
$z^\text{lab}=\cos\theta_\text{lab}$) and $z_\text{cut}^\text{lab} = \tanh
\eta_\text{cut} \simeq 0.987$.  These cuts should however be transformed from
the lab system to the c.m.\ system of the two leptons (where the leptons are
back-to-back). This gives for the cosine of the lepton angles in the c.m.\
system the following `visible' $z$ range
\begin{alignat}{3}
-\tanh(\eta_\text{cut} + y) 
&< \phantom{-}z &< \tanh(\eta_\text{cut} - y),
\quad &\text{for } l^-,
\nonumber \\
-\tanh(\eta_\text{cut} + y) 
&< - z &< \tanh(\eta_\text{cut} - y),
\quad &\text{for } l^+,
\end{alignat}
where $y$ is the rapidity of the c.m.\ frame of the lepton pair w.r.t.\ the
lab frame ($pp$ c.m.\ frame).  Note that for a given $y$ the visible
$z$ region for detection of one lepton is not symmetric around
$z=0$ unless $y=0$.  Since we require {\it both} leptons to be
detected, we combine the two regions given above into the symmetric region
\begin{align} \label{Eq:lhc-cuts}
|z| < \tanh(\eta_\text{cut} - |y|) = \frac{z_\text{cut}^\text{lab}
-|\beta|}{1-z_\text{cut}^\text{lab} |\beta|} \equiv z_\text{cut},
\end{align}
where $\beta=\tanh y$.  This means that we have to require
$|y|<\eta_\text{cut}=2.5$ in addition to the limits $|y|<Y=\log(\sqrt{s}/M)$,
hence $y_\text{max}=\min(\eta_\text{cut},Y)$. At the LHC, the cut
$|y|<\eta_\text{cut}$ affects the rapidity integration range for values of $M$
below $1.15$~TeV.

In addition to the angular cuts, we impose on each lepton a transverse
momentum cut
\begin{equation}
\label{Eq:pperp-cut}
p_\perp>p_\perp^\text{cut}=20\text{ GeV},
\end{equation}
which leads to $|z|<\sqrt{1-(2p_\perp^\text{cut}/M)^2}$. This cut affects the
$z$ integration range for $M<0.25$~TeV at the LHC.
Thus, the over-all cut on $z$ will in general depend on both $y$ and $M$.
\subsection{Tevatron}\label{subsec:cut-tevatron}
At the Tevatron, the cuts are more complicated, since both detectors there
have some additional coverage close to the beam pipe (`end plugs') in addition
to the central part of the detector. As an example, we shall here consider the
following cuts. We want either both leptons in the Central Calorimeter (CC),
$|\eta|< 1.1$ (corresponding to $|z^\text{lab}|< 0.800$) or one lepton in the
Central Calorimeter and one lepton in an End Cap (EC), $1.5 < |\eta| < 2.5$
(corresponding to $0.905 < |z^\text{lab}| < 0.987$)
\cite{Abachi:1996hs,Giordani:2003ib} (see Fig.~\ref{Fig:tevatron-cuts}).  We
get the restrictions on the $z$ range given in Appendix~A, and shown in
Fig.~\ref{Fig:tevatron-zcuts}.

\begin{figure}[htb]
\refstepcounter{figure}
\label{Fig:tevatron-cuts}
\addtocounter{figure}{-1}
\begin{center}
\setlength{\unitlength}{1cm}
\begin{picture}(7.0,7.0)
\put(0.0,0.5)
{\mbox{\epsfysize=6.5cm\epsffile{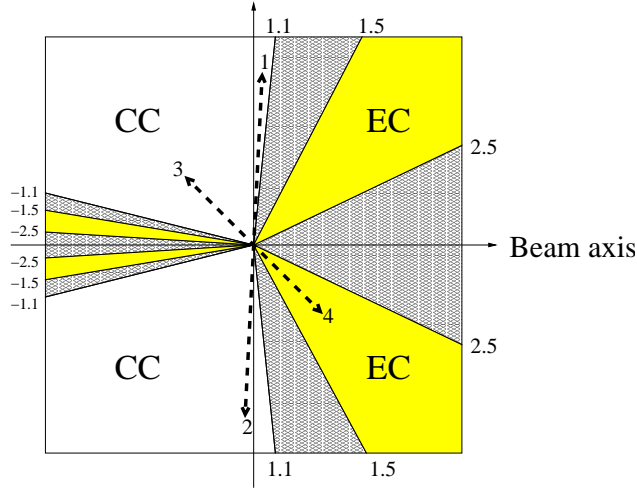}}}
\end{picture}
\vspace*{-8mm}
\caption{Schematic side view of Tevatron detector, as seen in the c.m.\ frame
at a positive boost given by $y=1.0$ ($\beta=0.762$): Central Calorimeter (CC)
and End Caps (EC) are indicated as white and yellow (shaded), with `blind'
directions dark.  The borders between different regions are labeled by the
corresponding pseudorapidity, $|\eta|=1.1$, 1.5 or 2.5.  The tracks labeled
`1' and `2' are back-to-back in the CC, whereas `3' and `4' are back to back
with `3' in the CC and `4' in the EC, compare Table~\ref{tab:zcuts-tevatron}
in Appendix~A.}
\end{center}
\end{figure}

\begin{figure}[htb]
\refstepcounter{figure}
\label{Fig:tevatron-zcuts}
\addtocounter{figure}{-1}
\begin{center}
\setlength{\unitlength}{1cm}
\begin{picture}(7.0,7.0)
\put(0.0,0.5)
{\mbox{\epsfysize=7.5cm\epsffile{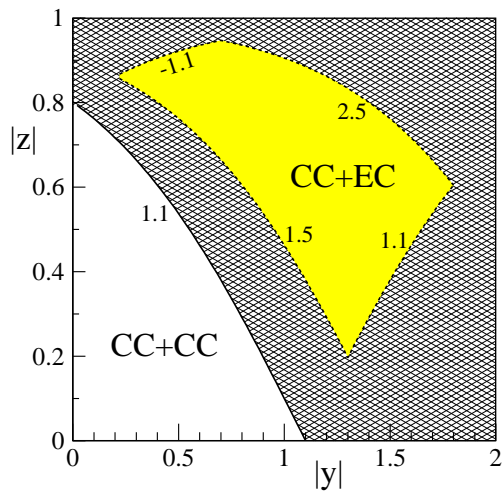}}}
\end{picture}
\vspace*{-8mm}
\caption{Cuts on $z=\cos\theta_\text{cm}$ vs.\ c.m.\
rapidity $y$ for the Tevatron.  White: region allowed by having both
leptons in the Central Calorimeter; yellow (shaded): region
allowed by having one lepton in the Central Calorimeter and the other
in an End Cap. Cuts are labeled by the corresponding pseudorapidities,
compare Table~\ref{tab:zcuts-tevatron} in Appendix~A.}
\end{center}
\end{figure}

Since for rapidity $y$ larger than 1.8 there is no integration range left,
these cuts result in a rapidity cut in addition to the limit
$|y|\le\log(\sqrt{s}/M)$.  This additional limitation, $|y|\le1.8$, only
affects the integration range in $y$ for invariant dilepton masses
$M<0.32$~TeV at the Tevatron.  Like for the LHC, we also impose the $p_\perp$
cut of Eq.~(\ref{Eq:pperp-cut}).

Note that terms that are odd in $z=\cos\theta_\text{cm}$ do not contribute to
$A_\text{CE}$, since, as previously emphasized, the integration limits are
always symmetric around $z=0$. Consequently, the applied experimental cuts in
the laboratory frame must respect this symmetry in order to measure this
observable.

\subsection{The vanishing of $A_\text{CE}^\text{SM}$}
\label{subsect:vanishing}
As pointed out in the final part of Sec.~\ref{subsec:add}, the zero point of
$A_\text{CE}$ can be a good indicator of the usefulness of this observable in
discriminating against the SM as well as any new physics based on vector
couplings.
If the zero point is little changed by the inclusion of gravity
effects, then it is unlikely that $A_\text{CE}$ will be useful.  
We will now discuss approaches for obtaining $A_\text{CE}^\text{SM}=0$
in the presence of angular cuts.
This can be done either differentially in $y$, or after an integration
over $y$.
\subsubsection{A rapidity-dependent approach}
One can define a value $\hat z_0^*$, such that the parton-level center--edge
cross section vanishes:
\begin{equation} \label{Eq:sigma-hat-SM=0}
\hat\sigma_\text{CE}^\text{SM}(z^*=\hat z_0^*)=0.
\end{equation}
In the case of no angular cuts, $\hat z_0^*=z_0^*$, as given by
Eq.~(\ref{Eq:z0*}). However, as anticipated at the beginning of this section,
the introduction of cuts will contribute to a modification of this value,
which will depend on both the rapidity and the invariant dilepton mass $M$
considered.

When angular cuts are introduced ($|z|\le z_\text{cut}$),
it follows from Eq.~(\ref{Eq:sigma_ce}) that
\begin{equation} \label{Eq:sigma-hat-SM}
\hat\sigma_\text{CE}^\text{SM}(z^*)
\propto z^*(z^{*2}+3)-\half\, z_\text{cut}[z_\text{cut}^2 + 3].
\end{equation} 
One can thus determine the value of $z^*$ for which
$\hat\sigma_\text{CE}^\text{SM}$ vanishes, by solving a cubic equation.  
The solution can for the LHC be given by the simple expression
\begin{equation}
\label{Eq:z0star-y}
\hat z_0^*=(\sqrt{b}+a)^{1/3}-(\sqrt{b}-a)^{1/3},
\end{equation}
where
\begin{equation}
a=\fourth z_\text{cut}[z_\text{cut}^2+3], \quad
b=a^2+1,
\end{equation}
generalizing Eq.~(\ref{Eq:z0*}).  In Fig.~\ref{Fig:z0-y-dependence}, we
display both $z_\text{cut}$ and $\hat z_0^*$ as functions of $y$ for two
values of the invariant dilepton mass. Since $z_\text{cut}=z_\text{cut}(y)$ as
given in Sec.~\ref{subsec:cut-lhc}, $\hat z_0^*$ will also depend on $y$, but
not on $M$ for $M>0.25~\text{TeV}$, where the $p_\perp$ cut of
Eq.~(\ref{Eq:pperp-cut}) plays no role.

At the Tevatron, the corresponding cubic equation is more complicated, since
there for some range of $|y|$ are two disjoint regions of $|z|$, and at large
$|y|$, a region where low $|z|$ are disallowed (see
Fig.~\ref{Fig:tevatron-zcuts}).
\begin{figure}[htb]
\refstepcounter{figure}
\label{Fig:z0-y-dependence}
\addtocounter{figure}{-1}
\begin{center}
\setlength{\unitlength}{1cm}
\begin{picture}(16.,7.7)
\put(0.0,0.0)
{\mbox{\epsfysize=8.0cm\epsffile{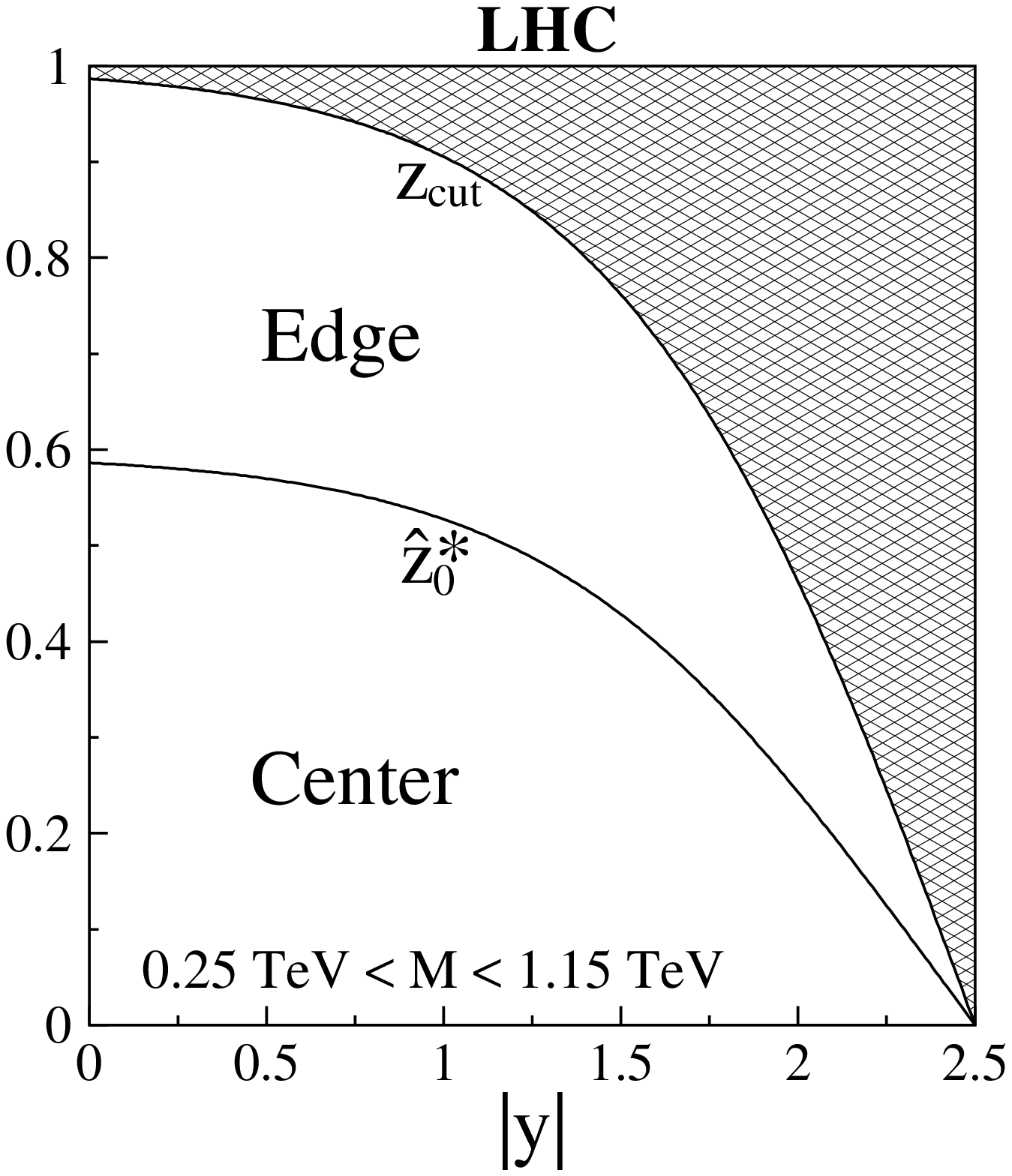}}
 \mbox{\epsfysize=8.0cm\epsffile{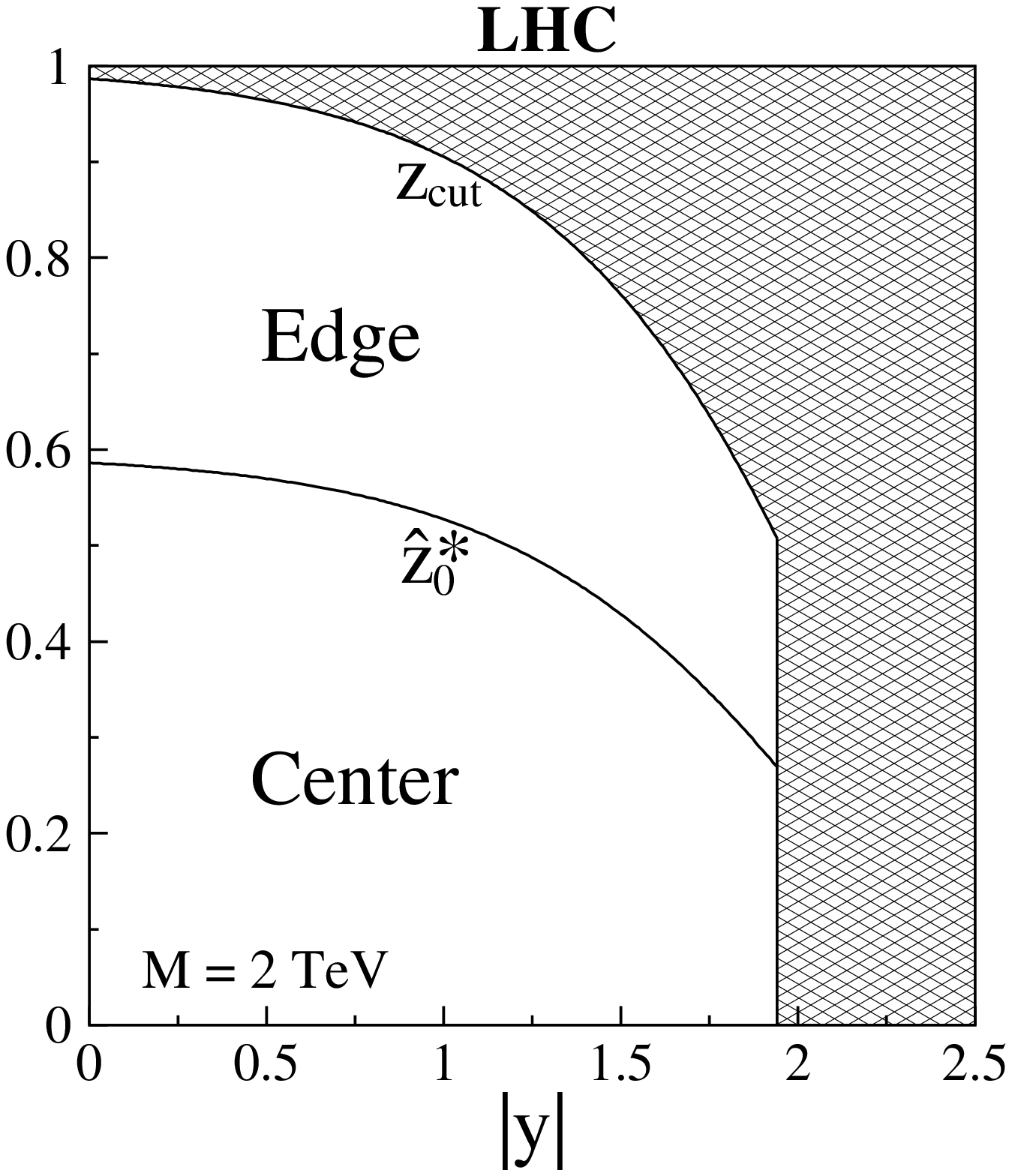}}}
\end{picture}
\vspace*{-8mm}
\caption{Dependence of $z_\text{cut}$ and $\hat z_0^*$ on $|y|$ at the LHC for
$0.25~\text{TeV}<M<1.15~\text{TeV}$ and $M=2$~TeV. The dark regions are
excluded by the cuts,  the additional limitation on $|y|$ in the right-hand
panel is due to the $M$-dependent kinematical limit $|y|\le Y$.}
\end{center}
\end{figure}

\subsubsection{Zeros of $A_\text{CE}$ and their $M$ dependence}
\label{subsect:z0star}
It is also possible to make $A_\text{CE}^\text{SM}$ vanish after the
introduction of cuts, by redefining $z_0^*$ such that
\begin{equation}
\frac{d\sigma_\text{CE}^\text{SM}}{dM}(z^*=z_0^*)=0.
\end{equation}
This provides a more global value for $z_0^*$, but since the parton
distribution functions are involved in the integration over $y$, we cannot
find an analytic expression for this quantity.

With the cuts discussed in Secs.~\ref{subsec:cut-lhc} and
\ref{subsec:cut-tevatron}, and using the center--edge definition in
Eq.~(\ref{Eq:ace}), $z_0^*$ depends on the invariant mass of the lepton-pair
as shown in Fig~\ref{Fig:z0-hats-dependence} (the SM curve is the same in both
panels).  In addition to $z_0^*$ (solid), we also plot the contours
corresponding to $A_\text{CE}=0$ for $\lambda=\pm1$ (dashed) in the $z^*$--$M$
plane for two cases, $M_H=2$ and $4$~TeV, both at the LHC.
\begin{figure}[htb]
\refstepcounter{figure}
\label{Fig:z0-hats-dependence}
\addtocounter{figure}{-1}
\begin{center}
\setlength{\unitlength}{1cm}
\begin{picture}(16.2,7.7)
\put(0.0,0.0)
{\mbox{\epsfysize=8.0cm\epsffile{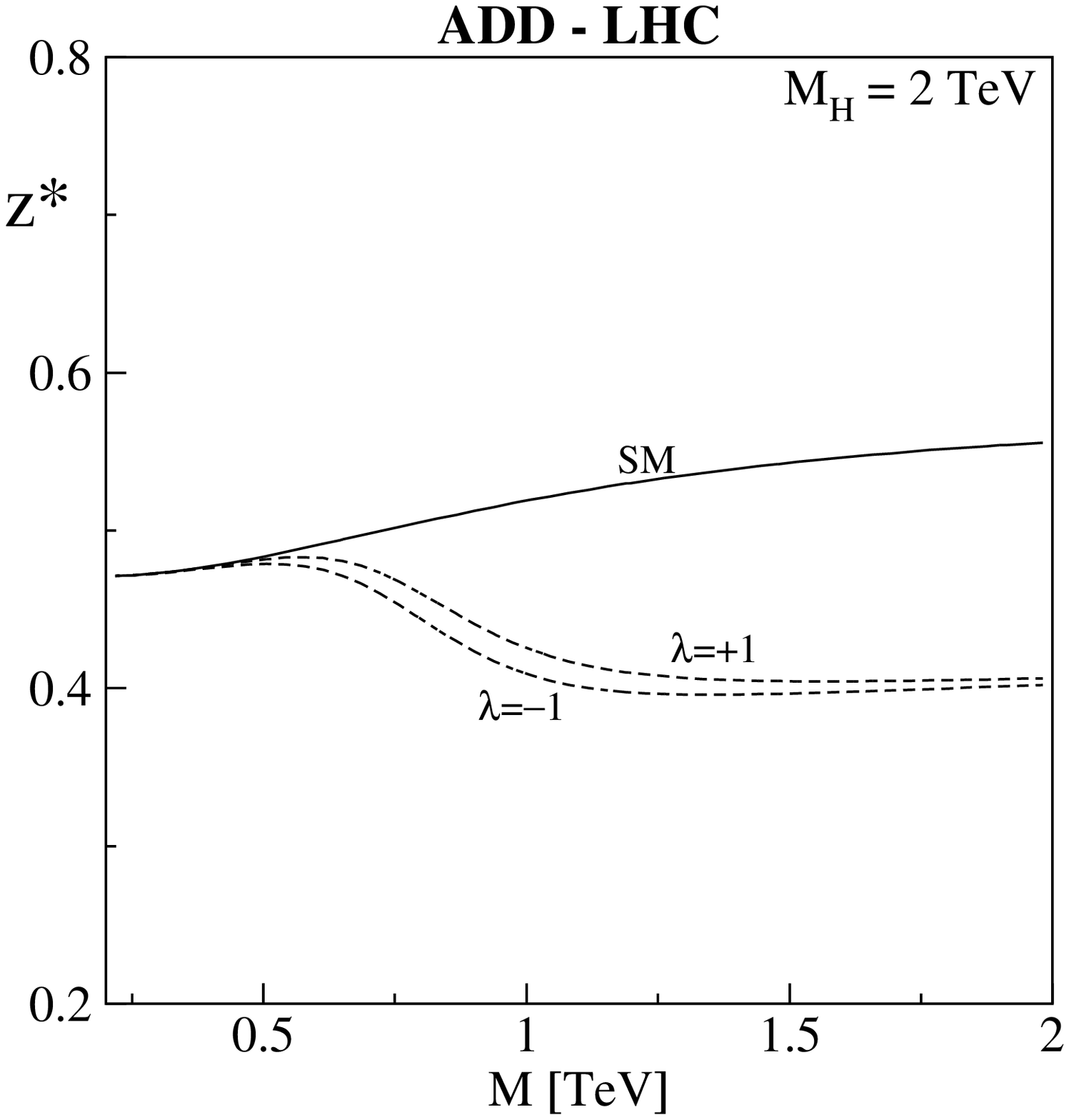}}
 \mbox{\epsfysize=8.0cm\epsffile{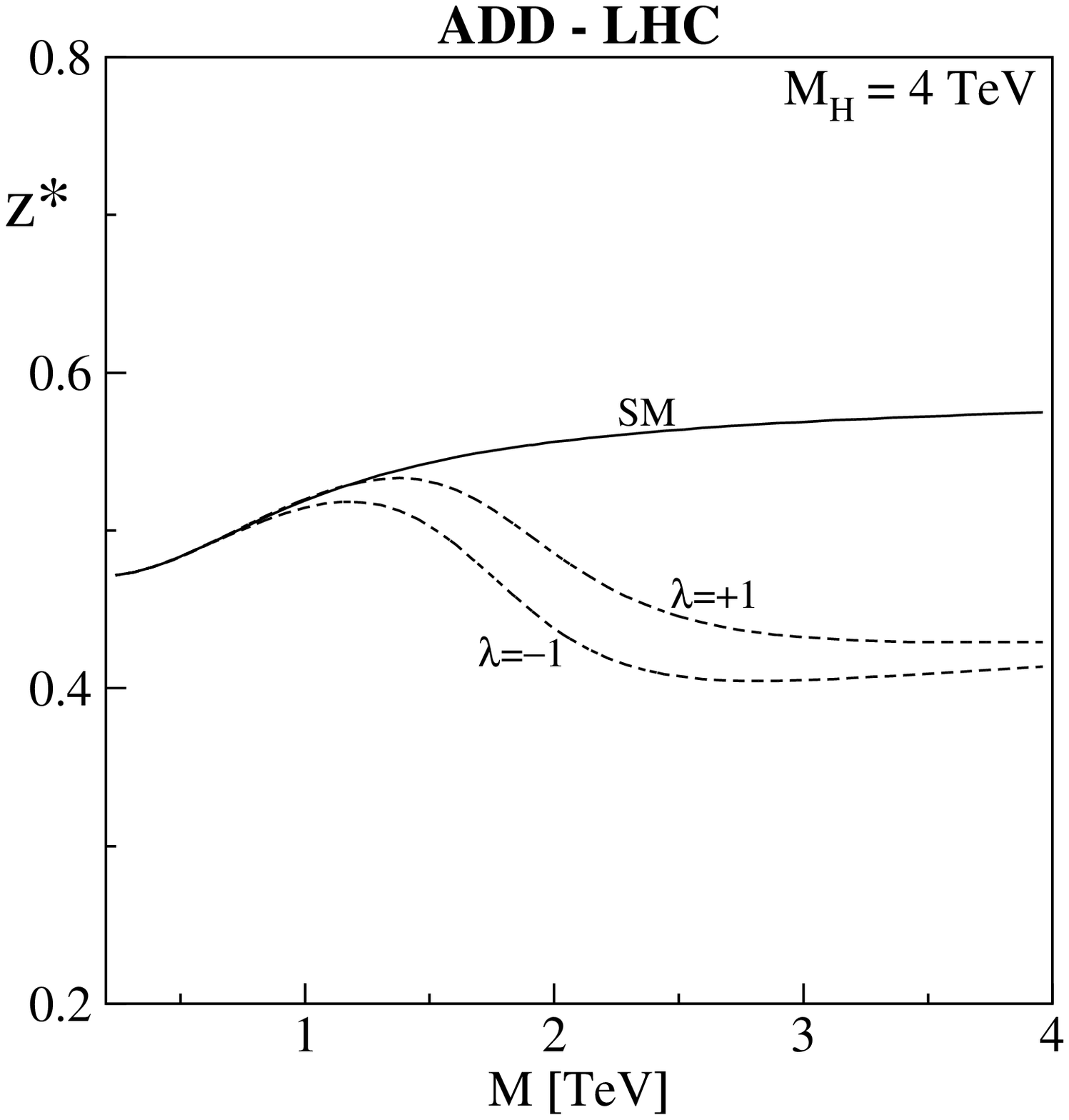}}}
\end{picture}
\vspace*{-8mm}
\caption{Zeros of $A_\text{CE}$ and their $M$ dependence at the LHC.  The
solid curves give $z_0^*$, which corresponds to $A_\text{CE}^\text{SM}=0$,
whereas the dashed curves correspond to $A_\text{CE}=0$ for the ADD model,
with $M_H=2$ and $4$~TeV.}
\end{center}
\end{figure}

The $M$-dependence of $z_0^*$ can qualitatively be understood as follows.  At
low invariant masses, also high $y$ contribute to the cross section, but here
$z$ has an upper bound significantly below 1 (see
Fig.~\ref{Fig:z0-y-dependence}). For higher masses, on the other hand, the
dominant contribution comes from lower values of $y$, where $z$ can reach
higher values. In order to keep $d\sigma_\text{CE}^\text{SM}/dM=0$, $z_0^*$
has to increase with increasing mass $M$ and tends to its asymptotic value
$\approx 0.586$, as seen in Fig.~\ref{Fig:z0-hats-dependence}.

Let us next consider the difference between $z_0^*$ for the SM and the $z^*$
value at which $A_\text{CE}$ vanishes for the ADD case.  First of all, at low
$M$, graviton exchange does not play any role, due to the strongly suppressing
factors $(M/M_H)^4$ and $(M/M_H)^8$, see Eq.~(\ref{Eq:sigma_ce}), so the SM
and ADD curves coincide.  At the LHC, where pure graviton exchange gives an
important contribution to $A_\text{CE}$, $z_0^*$ is larger, since pure
graviton events are more central, as given by Eq.~(\ref{Eq:parton-CE-limits}).

At the Tevatron, the more complicated cuts lead to a different behavior, with
decreasing $z_0^*$ as $M$ increases.  In addition, the values for $z_0^*$ for
the SM and the ADD model are rather close in the region where the majority of
events will occur. Consequently, in the case of the Tevatron, the usefulness
of $A_\text{CE}$ is in fact not a priori guaranteed.

\section{Identification of spin-2}
\label{sec:identify}
\setcounter{equation}{0}
In this section we assume that a deviation from the SM is discovered in the
cross section, either in the form of a contact interaction or a resonance. We
will here investigate in which regions of the ADD and RS parameter spaces such
a deviation can be identified as being caused by spin-2 exchange.  More
precisely, we will see how the center--edge asymmetry can be used to exclude
spin-1 exchange beyond that of the SM.

In order to get more statistics, one may integrate over bins $i$ in $M$. We
therefore define the bin-integrated center--edge asymmetry by introducing such
an integration,
\begin{equation}
\label{Eq:acecut}
A_\text{CE}(i)
=\frac{\displaystyle{\int_i \frac{d\sigma_\text{CE}}{dM}}dM}
{\displaystyle{\int_i \frac{d\sigma}{dM}dM}}.
\end{equation}
Here, the rapidity-dependent approach,
with $\hat z_0^*(y)$ given by Eq.~(\ref{Eq:z0star-y}), and shown in
Fig.~\ref{Fig:z0-y-dependence} for the LHC, will be used.  The advantage of
this approach is that it is given by an explicit analytical expression,
whereas the approach outlined in subsec.~\ref{subsect:z0star} depends on the
$M$-range considered.

To determine the underlying new physics (spin-1 vs.\ spin-2 couplings) one can
introduce the deviation of the measured center--edge asymmetry from that
expected from pure spin-1 exchange, $A_\text{CE}^\text{spin-1}(i)$ (which in
our approach is zero), in each bin,
\begin{equation}  \label{Eq:Delta-ACE}
\Delta A_\text{CE}(i)=A_\text{CE}(i)-A_\text{CE}^\text{spin-1}(i).
\end{equation}
The bin-integrated statistical uncertainty is then given as
\begin{equation} \label{Eq:fiveone}
\delta A_\text{CE}(i)=\sqrt{\frac{1-A_\text{CE}^2(i)}
{\epsilon_l {\cal L}_\text{int}\sigma(i)}}, 
\end{equation}
based on the number of events that are effectively detected
and the $A_\text{CE}$ that is actually measured.  
We take the efficiency for reconstruction of lepton pairs, $\epsilon_l=90\%$
and sum over $l=e,\mu$.

The statistical significance, ${\cal S}_\text{CE}(i)$ is defined as:
\begin{equation} \label{Eq:stat-sign-add}
{\cal S}_\text{CE}(i)=\frac{|\Delta A_\text{CE}(i)|}
{\delta A_\text{CE}(i)}.
\end{equation}

\subsection{ADD case}
In the ADD scenario, the identification reach in $M_H$ can
be estimated from a $\chi^2$ analysis:
\begin{equation} \label{Eq:five-three}
\chi^2=\sum_i\left[{\cal S}_\text{CE}(i)\right]^2,
\end{equation}
where $i$ runs over the different bins in $M$.
The 95\% CL is then obtained by requiring $\chi^2=3.84$, as pertinent
to a one-parameter fit \cite{eadie}.

At the LHC, with $100~\text{fb}^{-1}$, we require $M>400$~GeV and divide the
data into $200$~GeV bins as long as the number of events in each bin,
$\epsilon_l {\cal L}_\text{int}\sigma(i)$, is larger than 10.  Therefore, the
number of bins will depend on the 
magnitude of the (discovered) deviation from the SM.

We find that at the $95 \%$ CL, the {\it identification} reach at the LHC,
where one can distinguish between the ADD and an alternative spin-1 based
scenario, is $M_H = 4.77$~TeV and $5.01$~TeV for $\lambda=+1$ and $-1$,
respectively. In the first case, we used $17$ bins, whereas in the last case
the number of bins was $13$ (the numbers of bins are determined
by the condition of having at least 10 events in each bin).
If no cuts are imposed, one would obtain an improvement of the
identification reach on $M_H$, by up to 2\%, with
$z_0^*\simeq0.596$. Therefore, at the LHC, the cuts have only a moderate
impact on the results.

At the Tevatron, with $2~\text{fb}^{-1}$, the lower limit on $M$ is chosen to
be $M>200$~GeV, with $50$~GeV bins.  We find for the identification reach,
$M_H = 0.87$~TeV and $0.97$~TeV for $\lambda=+1$ and $-1$, with 13 and 8 bins,
respectively.  If no cuts are imposed, one would obtain slightly improved
limits, namely $M_H = 0.99$~TeV and $1.06$~TeV for the two cases, with 10 and
8 bins and $z_0^*\simeq0.596$.  Thus, the ``modest'' identification reach at
the Tevatron is due to a combination of several effects (including partial
cancellation of interference related to $u$ and $d$ quarks), and not primarily
determined by the more severe cuts. However, these regions of parameter space
have already been excluded by LEP \cite{geweniger} which has reached
$M_H=1.20$ and 1.09~TeV (for $\lambda=+1$ and $-1$) (for Fermilab results, see
\cite{Abbott:2000zb,Sanders:2003ye}).
\begin{table}[ht]
\begin{center}
\caption{Identification reach on $M_H$ (in TeV) at 95\% CL 
from $A_\text{CE}$.}
\label{tab:identify}
\vspace{.175in}
\renewcommand{\tabcolsep}{.75em}
\begin{tabular}{|c|c|c|c|}
\hline 
Collider&$\lambda=+1$&$\lambda=-1$\\
\hline
Tevatron $2~\text{fb}^{-1}$& 0.9 & 1.0 \\
LHC $100~\text{fb}^{-1}$ & 4.8 & 5.0 \\
LHC $300~\text{fb}^{-1}$ & 5.4 & 5.9 \\
\hline
\end{tabular} 
\end{center}
\end{table}

In Table~\ref{tab:identify} we summarize the results, and also include the
identification reach corresponding to an integrated luminosity of
$300~\text{fb}^{-1}$ at the LHC.  The obtained results depend on the procedure
adopted for the binning.  Let us first discuss the case when we keep the range
in $M$ fixed, $M_\text{min}\le M \le M_\text{max}$, but increase the bin
width.  The identification reach turns out to be rather independent of the bin
width, as long as it does not increase by more than a factor of order two,
compared to the value we have adopted.  The reason is that when the bins
become too large, then contributions from low masses, where there are many
(SM) events, will dilute the signal in $A_\text{CE}$.  On the other hand, if
we make the bin width smaller, we would have to reduce $M_\text{max}$, in
order to satisfy the condition of having at least 10 events in the highest
bin. Such a reduction of $M_\text{max}$ normally leads to reduced sensitivity,
since the new physics effects are strongest at high masses.

Consider next the case when we relax the constraint of keeping $M_\text{max}$
fixed. Then, by increasing the chosen bin width, one could obtain a somewhat
higher identification reach on $M_H$, since (in the approach we follow here)
larger bin width would allow us to increase $M_\text{max}$. 
However, keeping in mind the ``effective character" of the interaction (\ref{Eq:Hewett}), 
such that  $M_H$ represents integration over
masses up to a cut-off, one should be careful not to go too close to $M_H$.

In a less conservative approach, one could put all events between
$M_\text{max}$ and $0.9\times M_H$ into one additional bin.  If one were to do
that, the identification reach would increase.  For example, for the LHC case,
with $100~\text{fb}^{-1}$ (see Table~\ref{tab:identify}), the identification
reach would increase by 4\% and 13\% for $\lambda=+1$ and $-1$, respectively.
\subsection{RS case}
A very distinct feature of the RS scenario is that the resonances are unevenly
spaced. If the first resonance is sufficiently heavy, the second resonance
would be difficult to resolve within the kinematical range allowed
experimentally.  For $m_1>1.7$, $2.5$, $2.8$~TeV for $k/\bar
M_\text{Pl}=0.01$, $0.05$, $0.1$, respectively, the second resonance would
contain less than 10 events at the LHC, for $100~\text{fb}^{-1}$ (in the
narrow-width approximation).  The corresponding critical values at the
Tevatron, for 2~$\text{fb}^{-1}$, are $m_1>0.28$, $0.44$, $0.50$~TeV.  In this
situation it would be of crucial importance to have a method of distinguishing
between spin-1 and spin-2 resonances and, indeed, this is what the
center--edge asymmetry can offer.

At the LHC (Tevatron), we choose a 200~GeV ($50$~GeV) bin around the resonance
mass $m_1$, and obtain the results presented in Fig.~\ref{Fig:rs-reach}, where
we display the 2, 3 and $5\sigma$ contours, cf.\ Eq.~(\ref{Eq:stat-sign-add}),
specialized to a single bin around $m_1$.  In order not to create additional
hierarchies, we require the scale of physical processes on the `TeV brane',
$\Lambda_\pi=m_1/[x_1(k/\bar M_\text{Pl})]<10~\text{TeV}$, as indicated in the
figure \cite{Davoudiasl:2000jd}.  Note that a spin-1 resonance (e.g., a $Z'$)
would give $\Delta A_\text{CE}=0$, provided we use $\hat z_0^*$ as given by
Eq.~(\ref{Eq:z0star-y}), or its analogue for the Tevatron.

\begin{figure}[htb]
\refstepcounter{figure}
\label{Fig:rs-reach}
\addtocounter{figure}{-1}
\begin{center}
\setlength{\unitlength}{1cm}
\begin{picture}(16.2,7.7)
\put(0.0,0.0)
{\mbox{\epsfysize=8.0cm\epsffile{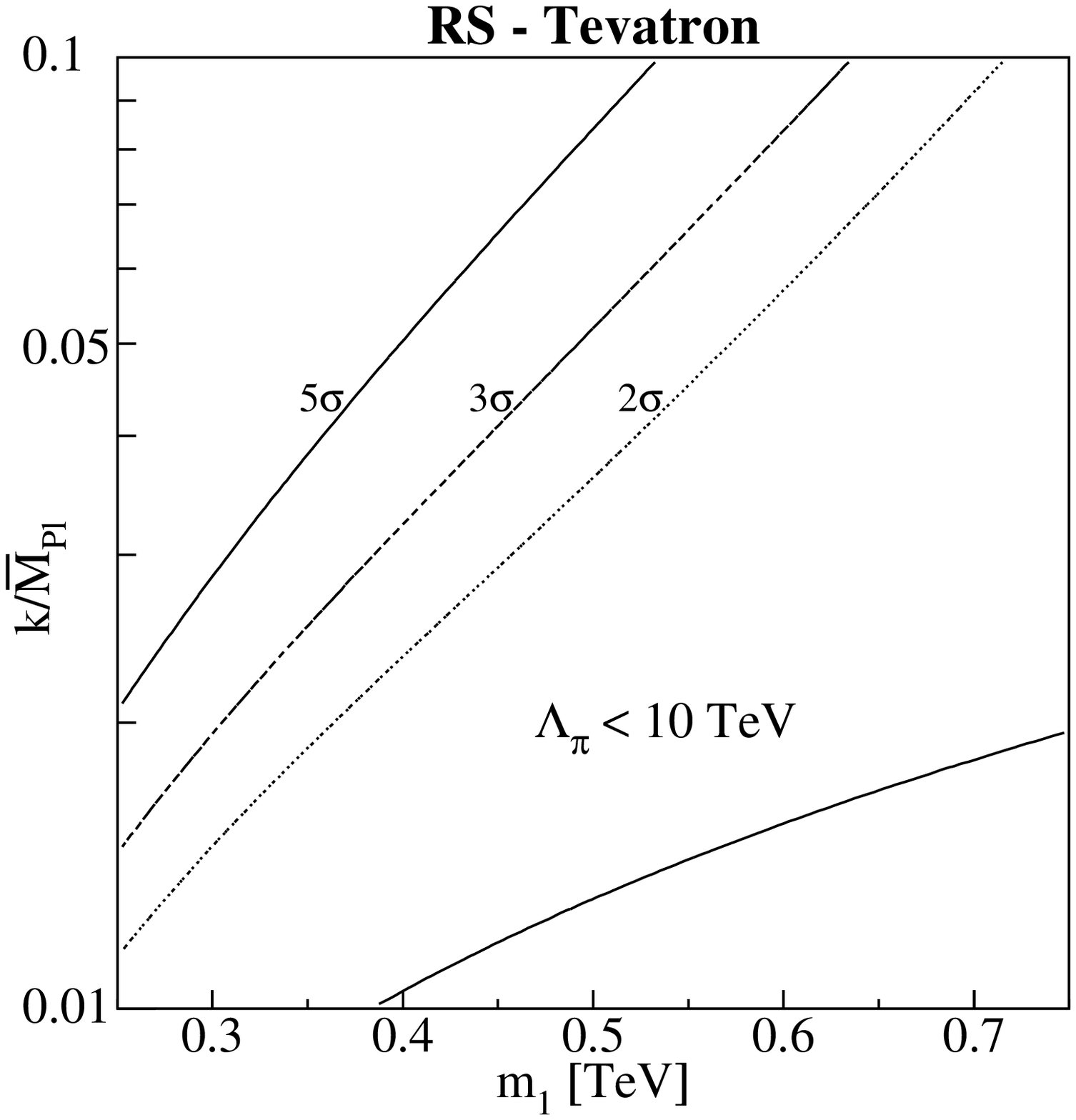}}
 \mbox{\epsfysize=8.0cm\epsffile{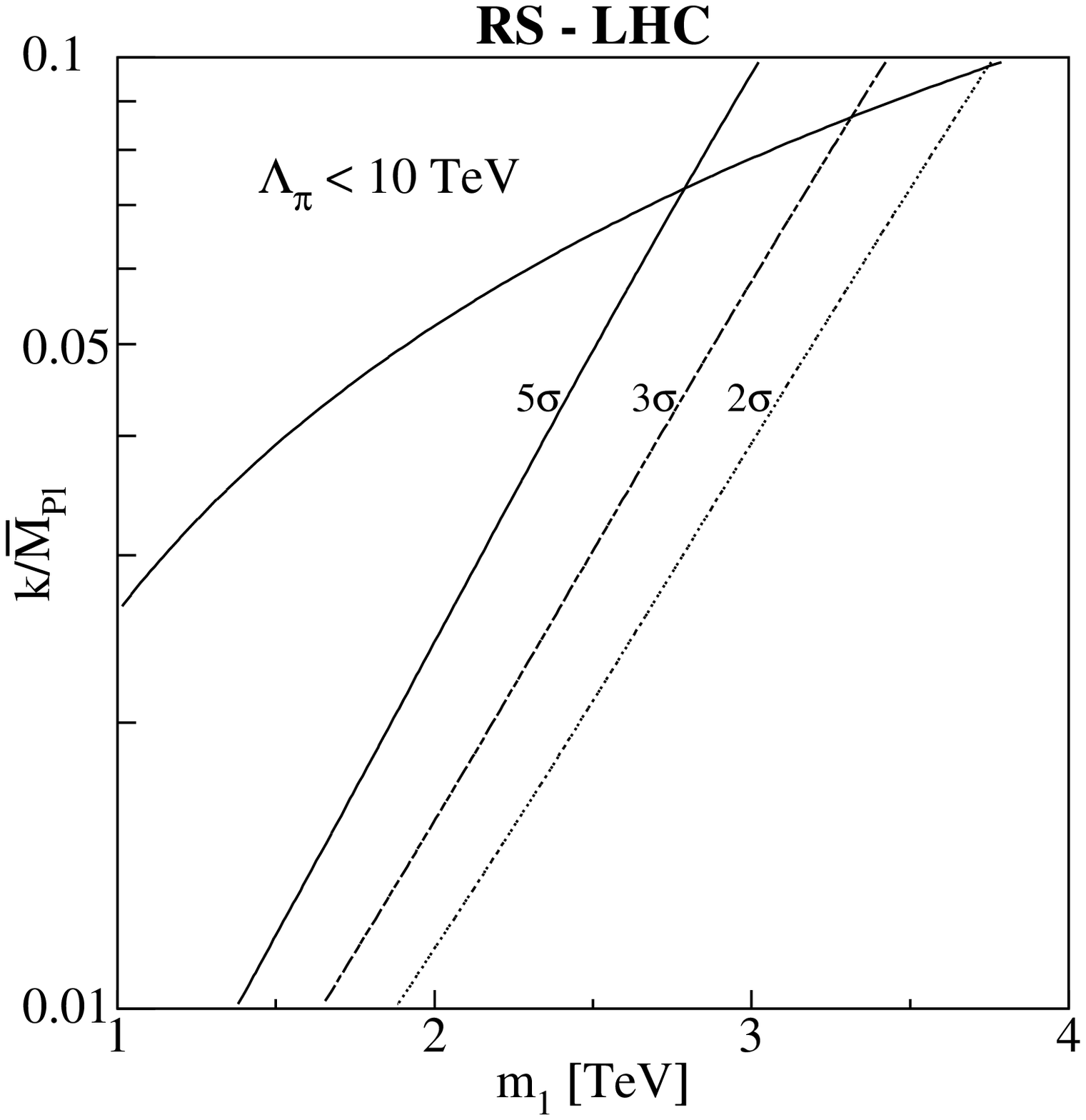}}}
\end{picture}
\vspace*{-8mm}
\caption{Spin-2 identification of an RS resonance, using the center--edge
asymmetry. We integrate over bins of 50 and 200~GeV around the peak at the
Tevatron and LHC, respectively. The theoretically favored region,
$\Lambda_\pi<10~\text{TeV}$, is indicated.}
\end{center}
\end{figure}

In the RS scenario, the LHC is capable of excluding or discovering graviton
resonances \cite{Traczyk:2002jh} in the whole region referred to as the
theoretically preferred part of the parameter space ($\Lambda_\pi<{\cal
O}(10~\text{TeV})$). We see that the center--edge asymmetry can distinguish
spin-2 exchange from spin-1 exchange, in large parts of this
theoretically favoured region.  Note also that the corresponding bounds cover
the regions of parameter space where only the first resonance is produced.
The position of a second resonance would reveal its RS-nature from the mass
splitting.

\section{Concluding remarks}\label{sec:summary}
\setcounter{equation}{0}

Exploring the center--edge asymmetry at hadron colliders is a good strategy to
distinguish between spin-1 and spin-2 exchange. 
The proposed center-edge asymmetry may be seen as a possible alternative or
supplement to a direct fit to the differential angular distribution
\cite{Allanach:2000nr,Traczyk:2002jh}. 

We have considered the ADD scenario parametrized by $M_H$, and the RS scenario
parametrized by $m_1$ and $k/\bar M_\text{Pl}$.
Although somewhat higher sensitivity reaches on $M_H$ or $m_1$ than obtained here
are given by other approaches, this method based on
$A_\text{CE}$ is suitable for actually pinning down the
spin-2 nature of the KK gravitons up to very high $M_H$ or $m_1$. This is
different from just detecting deviations from the Standard Model
predictions, and is a way
to obtain additional information on the underlying new-physics scenario.

The results obtained appear to be only moderately dependent on the
parametrization of the parton distribution functions used. Compared to CTEQ6
\cite{Pumplin:2002vw}, the MRST parametrization \cite{Martin:1998np}, gives
practically no change in $A_\text{CE}$ (since this is obtained as a ratio),
but changes the identification reach by 1-2\%, since the cross section
increases and therefore the statistics improve.

The {\it identification} of a spin-2 exchange relies on first discovering a
deviation from the Standard Model. For this purpose, the conventional cross
section, $\sigma$, is more sensitive than $A_\text{CE}$.  The sensitivity of
$\sigma$ can be further extended by considering also the forward--backward
asymmetry, $A_\text{FB}$, especially at the Tevatron, where the $q\bar q$
channel is important. At the LHC, $A_\text{FB}$ is less useful, since it is
insensitive to the gluon--gluon channel, as already remarked in
Sec.~\ref{sec:no-cuts}, and since a cut of $|y|>1$ is needed to enhance the
probability to experimentally identify the proton from which the quark
originated, thus reducing the number of events.

At a hadron collider, the case of spin-0 exchange would be more difficult to
discriminate against, since the difference in $A_\text{CE}$ would be
smaller. However, at an electron-positron collider with polarized beams, there
are more observables available, and hence a discrimination might be possible
\cite{Rizzo:1998vf,Osland:2003fn}.

We have here considered the dilepton channels.
The identification reach can of course be extended
by the inclusion of the diphoton and possibly also the dijet channels.

\bigskip
\medskip
\leftline{\bf Acknowledgment}
\par\noindent
We are grateful to Tom Rizzo for useful communication.
One of us (AAP) wants to express his gratitude to N.\ Rusakovich and
N.\ Shumeiko for stimulating discussions.
This research has been partially supported by MIUR (Italian Ministry of
University and Research), by funds of the University of Trieste,
and by the Research Council of Norway.

\appendix
\renewcommand{\theequation}{\thesection\arabic{equation}}

\section*{Appendix A}
\setcounter{equation}{0}
\renewcommand{\thesection}{A}

In order to provide a compact description of the integration ranges
of $z$ that are involved for various values of the rapidity
$y$, let us introduce the abbreviation
\begin{equation}
z_\text{cut}(\eta_i,y)
=\tanh(\eta_i - |y|).
\end{equation}
According to Sec.~\ref{subsec:cut-tevatron}, we determine the cuts in $z$, for
$\eta_1=1.1$, $\eta_2=1.5$ and $\eta_3=2.5$.  Leptons are either both in the
Central Calorimeter, $|\eta|<1.1$ (``CC \& CC'') or one lepton is in the
Central Calorimeter and one is in the End Cap, $1.5<|\eta|<2.5$ (``CC \&
EC''). The resulting limits are summarized in Table~\ref{tab:zcuts-tevatron}.
\begin{table}[ht]
\begin{center}
\caption{Tevatron cuts on the c.m.\ angle, in terms of
$|z|$.}\label{tab:zcuts-tevatron}
\label{tab:cuts}
\vspace{.175in}
\renewcommand{\tabcolsep}{.75em}
\begin{tabular}{|c|c|c|c|}
\hline 
c.m.\ rapidity $y$& CC \& CC&CC \& EC\\
\hline
$0\le|y|\le0.2$&$|z| \le z_\text{cut}(1.1,y)$&\\
$0.2\le|y|\le0.7$&$|z| \le z_\text{cut}(1.1,y)$ &
          $z_\text{cut}(1.5,y) \le |z|\le 
           |z_\text{cut}(-1.1,y)|$ \\
$0.7\le|y|\le1.1$&$|z| \le z_\text{cut}(1.1,y)$ & 
          $z_\text{cut}(1.5,y) \le 
            |z|\le z_\text{cut}(2.5,y)$ \\
$1.1\le|y|\le1.3$& & 
          $z_\text{cut}(1.5,y) \le 
            |z|\le z_\text{cut}(2.5,y)$ \\
$1.3\le|y|\le1.8$& & 
          $|z_\text{cut}(1.1,y)| \le 
            |z|\le z_\text{cut}(2.5,y)$ \\
\hline
 \end{tabular} 
\end{center}
\end{table}

At low rapidities, there is only one range in $z$ (where both
leptons hit the CC), whereas for intermediate rapidities $y$ there are three
ranges: for `small' $|z|$ both leptons hit the CC, whereas at
`larger' values of $|z|$ ($z$ being positive or negative)
one lepton hits the CC and the other hits an EC (a representative case, with
$y=1$, is illustrated by Fig.~\ref{Fig:tevatron-cuts}).  Finally, at `large'
rapidities, there are only two ranges of positive or negative values of
$z$ for which one lepton hits the CC and the other hits an EC. 
For rapidities above $1.8$, there is no $z$ range left.

\goodbreak


\begin{thebibliography}{99}

\bibitem{'tHooft:xb}
G.~'t Hooft, in ``Recent Developments In Gauge Theories'', Proceedings, 
NATO Advanced Study Institute, Cargese, France, August 26 - September 8, 1979,
edited by
G.~'t Hooft, C.~Itzykson, A.~Jaffe, H.~Lehmann, P.~K.~Mitter, 
I.~M.~Singer and R.~Stora
{\it  New York, USA: Plenum (1980) 
(NATO Advanced Study Institutes Series: Series B, Physics, 59)}; \\
S.~Dimopoulos, S.~Raby and L.~Susskind,
Nucl.\ Phys.\ B {\bf 173} (1980) 208.

\bibitem{Eichten:1983hw}
E.~Eichten, K.~D.~Lane and M.~E.~Peskin,
Phys.\ Rev.\ Lett.\  {\bf 50} (1983) 811; \\
R.~Ruckl,
Phys.\ Lett.\ B {\bf 129} (1983) 363.

\bibitem{Barger:1997nf}
V.~D.~Barger, K.~m.~Cheung, K.~Hagiwara and D.~Zeppenfeld,
Phys.\ Rev.\ D {\bf 57} (1998) 391
[arXiv:hep-ph/9707412]; \\
D.~Zeppenfeld and K.~m.~Cheung,
Proceedings of 5th International WEIN Symposium: 
A Conference on Physics Beyond the Standard Model (WEIN 98), Santa Fe, NM, 
14--21 Jun 1998;
arXiv:hep-ph/9810277.

\bibitem{Hewett:1988xc}
For reviews see, {\it e.g.}:
J.~L.~Hewett and T.~G.~Rizzo,
Phys.\ Rept.\  {\bf 183} (1989) 193; \\ 
A.~Leike,
Phys.\ Rept.\  {\bf 317} (1999) 143
[arXiv:hep-ph/9805494];

\bibitem{Buchmuller:1986zs}
W.~Buchmuller, R.~Ruckl and D.~Wyler,
Phys.\ Lett.\ B {\bf 191} (1987) 442
[Erratum-ibid.\ B {\bf 448} (1999) 320];\\
G.~Altarelli, J.~R.~Ellis, G.~F.~Giudice, S.~Lola and M.~L.~Mangano,
Nucl.\ Phys.\ B {\bf 506} (1997) 3
[arXiv:hep-ph/9703276]; \\
R.~Casalbuoni, S.~De Curtis, D.~Dominici and R.~Gatto, 
Phys.\ Lett.\ B {\bf 460} (1999) 135
[arXiv:hep-ph/9905568];\\
V.~D.~Barger and K.~m.~Cheung,
Phys.\ Lett.\ B {\bf 480} (2000) 149
[arXiv:hep-ph/0002259]. 

\bibitem{Kalinowski:1997bc}
J.~Kalinowski, R.~Ruckl, H.~Spiesberger and P.~M.~Zerwas,
Phys.\ Lett.\ B {\bf 406} (1997) 314
[arXiv:hep-ph/9703436];
Phys.\ Lett.\ B {\bf 414} (1997) 297
[arXiv:hep-ph/9708272].

\bibitem{Rizzo:1998vf}
T.~G.~Rizzo,
Phys.\ Rev.\ D {\bf 59} (1999) 113004
[arXiv:hep-ph/9811440].

\bibitem{Gounaris:1997ft}
G.~J.~Gounaris, D.~T.~Papadamou and F.~M.~Renard,
Phys.\ Rev.\ D {\bf 56} (1997) 3970
[arXiv:hep-ph/9703281].

\bibitem{Hewett:1999sn}
J.~L.~Hewett,
Phys.\ Rev.\ Lett.\  {\bf 82} (1999) 4765
[arXiv:hep-ph/9811356].

\bibitem{Davoudiasl:2000jd}
H.~Davoudiasl, J.~L.~Hewett and T.~G.~Rizzo,
Phys.\ Rev.\ Lett.\  {\bf 84} (2000) 2080
[arXiv:hep-ph/9909255]; \\
%
H.~Davoudiasl, J.~L.~Hewett and T.~G.~Rizzo,
Phys.\ Rev.\ D {\bf 63} (2001) 075004
[arXiv:hep-ph/0006041].

\bibitem{Osland:2003fn}
P.~Osland, A.~A.~Pankov and N.~Paver,
Phys.\ Rev.\ D {\bf 68}, 015007 (2003)
[arXiv:hep-ph/0304123].

\bibitem{Antoniadis:1998ig}
I.~Antoniadis, N.~Arkani-Hamed, S.~Dimopoulos and G.~R.~Dvali,
Phys.\ Lett.\ B {\bf 436} (1998) 257
[arXiv:hep-ph/9804398]; \\
%
N.~Arkani-Hamed, S.~Dimopoulos and G.~R.~Dvali,
Phys.\ Rev.\ D {\bf 59} (1999) 086004
[arXiv:hep-ph/9807344]; \\
%
G.~F.~Giudice, R.~Rattazzi and J.~D.~Wells,
Nucl.\ Phys.\ B {\bf 544} (1999) 3
[arXiv:hep-ph/9811291]; 
%
Nucl.\ Phys.\ B {\bf 630} (2002) 293
[arXiv:hep-ph/0112161];\\
%
E.~A.~Mirabelli, M.~Perelstein and M.~E.~Peskin,
Phys.\ Rev.\ Lett.\  {\bf 82} (1999) 2236
[arXiv:hep-ph/9811337]; \\
%
T.~Han, J.~D.~Lykken and R.~J.~Zhang,
Phys.\ Rev.\ D {\bf 59} (1999) 105006
[arXiv:hep-ph/9811350];\\
%
K.~m.~Cheung,
Phys.\ Rev.\ D {\bf 61}, 015005 (2000)
[arXiv:hep-ph/9904266]; \\
%
L.~Randall and R.~Sundrum,
Phys.\ Rev.\ Lett.\  {\bf 83} (1999) 4690
[arXiv:hep-th/9906064];\\
%
S.~Cullen, M.~Perelstein and M.~E.~Peskin,
Phys.\ Rev.\ D {\bf 62} (2000) 055012
[arXiv:hep-ph/0001166]; \\
%
J.~Bijnens, P.~Eerola, M.~Maul, A.~M{\aa}nsson and T.~Sj{\"o}strand,
Phys.\ Lett.\ B {\bf 503} (2001) 341
[arXiv:hep-ph/0101316]; \\
%
T.~G.~Rizzo,
Phys.\ Rev.\ D {\bf 64} (2001) 095010
[arXiv:hep-ph/0106336];\\
%
K.~m.~Cheung and G.~Landsberg,
Phys.\ Rev.\ D {\bf 65} (2002) 076003
[arXiv:hep-ph/0110346];\\
%
G.~Pasztor and M.~Perelstein,
in {\it Proc. of the APS/DPF/DPB Summer Study on the Future of Particle 
Physics (Snowmass 2001) } ed. N.~Graf,
arXiv:hep-ph/0111471;\\
%
E.~Gabrielli and B.~Mele,
Nucl.\ Phys.\ B {\bf 647}, 319 (2002)
[arXiv:hep-ph/0205099]; \\
%
E.~Dvergsnes, P.~Osland and N.~\"Ozt\"urk,
Phys.\ Rev.\ D {\bf 67}, 074003 (2003)
[arXiv:hep-ph/0207221].

\bibitem{Allanach:2000nr}
B.~C.~Allanach, K.~Odagiri, M.~A.~Parker and B.~R.~Webber,
JHEP {\bf 0009}, 019 (2000)
[arXiv:hep-ph/0006114].

\bibitem{Arkani-Hamed:1998rs}
N.~Arkani-Hamed, S.~Dimopoulos and G.~R.~Dvali,
Phys.\ Lett.\ B {\bf 429} (1998) 263
[arXiv:hep-ph/9803315].

\bibitem{Randall:1999ee}
L.~Randall and R.~Sundrum,
Phys.\ Rev.\ Lett.\  {\bf 83} (1999) 3370
[arXiv:hep-ph/9905221].

\bibitem{Dienes:1998vh}
K.~R.~Dienes, E.~Dudas and T.~Gherghetta,
Phys.\ Lett.\ B {\bf 436}, 55 (1998)
[arXiv:hep-ph/9803466].

\bibitem{Rubakov:bb}
V.~A.~Rubakov and M.~E.~Shaposhnikov,
Phys.\ Lett.\ B {\bf 125} (1983) 136.

\bibitem{Antoniadis:1990ew}
I.~Antoniadis,
Phys.\ Lett.\ B {\bf 246} (1990) 377; \\
I.~Antoniadis and K.~Benakli,
Phys.\ Lett.\ B {\bf 326}, 69 (1994)
[arXiv:hep-th/9310151];
I.~Antoniadis, K.~Benakli and M.~Quiros,
Phys.\ Lett.\ B {\bf 331}, 313 (1994)
[arXiv:hep-ph/9403290].

\bibitem{Pankov:1997da}
A.~A.~Pankov, N.~Paver and C.~Verzegnassi,
Int.\ J.\ Mod.\ Phys.\ A {\bf 13}, 1629 (1998)
[arXiv:hep-ph/9701359].

\bibitem{Hewett:2002hv}
J.~Hewett and M.~Spiropulu,
Ann.\ Rev.\ Nucl.\ Part.\ Sci.\  {\bf 52}, 397 (2002)
[arXiv:hep-ph/0205106].

\bibitem{Giudice:2003tu}
G.~F.~Giudice and A.~Strumia,
Nucl.\ Phys.\ B {\bf 663}, 377 (2003)
[arXiv:hep-ph/0301232].

\bibitem{Gupta:1999iy}
A.~K.~Gupta, N.~K.~Mondal and S.~Raychaudhuri,
arXiv:hep-ph/9904234; \\
K.~m.~Cheung and G.~Landsberg,
Phys.\ Rev.\ D {\bf 62}, 076003 (2000)
[arXiv:hep-ph/9909218].

\bibitem{Datta:2002tk}
A.~Datta, E.~Gabrielli and B.~Mele,
Phys.\ Lett.\ B {\bf 552}, 237 (2003)
[arXiv:hep-ph/0210318].

\bibitem{Pumplin:2002vw}
J.~Pumplin, D.~R.~Stump, J.~Huston, H.~L.~Lai, P.~Nadolsky and W.~K.~Tung,
JHEP {\bf 0207} (2002) 012
[arXiv:hep-ph/0201195].

\bibitem{LHC-ref}
A. Airapetian {\it et al.}, 
{\em ATLAS detector and Physical Performance Technical Design Report,} \\
http://atlasinfo.cern.ch/Atlas/GROUPS/PHYSICS/TDR/access.html,\\
G. L. Bayatian {\it et al.}, {\em CMS Technical Proposal},
CERN LHCC 94-38 (1994), \\
http://cmsinfo.cern.ch/TP/TP.html.

\bibitem{Abachi:1996hs}
S.~Abachi  [The D0 Collaboration],
{\it The D0 upgrade: The detector and its physics},
FERMILAB-PUB-96-357-E, \\
http://library.fnal.gov/archive/test-preprint/fermilab-pub-96-357-e.shtml,\\
R. Blair et al [The CDF Collaboration] 
{\it The CDF II Detector, Technical Design Report},
FERMILAB-Pub-96/390-E, \\
http://www-cdf.fnal.gov/upgrades/tdr/tdr.html

\bibitem{Giordani:2003ib}
M.~P.~Giordani  [CDF and D0 Collaborations],
FERMILAB-CONF-03-348-E
{\it Presented at International Europhysics Conference on High-Energy 
Physics (HEP 2003), Aachen, Germany, 17-23 Jul 2003}; \\
L.~Cerrito  [the CDF Collaboration],
arXiv:hep-ex/0311050; \\
{}  [CDF Collaboration],
arXiv:hep-ex/0311039.

\bibitem{eadie}
W.~T. Eadie, D. Drijard, F.~E. James, M. Roos, B. Sadoulet,
{\it Statistical methods in experimental physics}
(American Elsevier, 1971).

\bibitem{geweniger} 
LEPEWWG $f\bar f$ SubGroup (C.~Geweniger {\it et. al.}),
Combination of the LEP II $f\bar f$ Results, 
CERN preprint LEP2FF/02-03 (October 2002).

\bibitem{Abbott:2000zb}
B.~Abbott {\it et al.}  [D0 Collaboration],
Phys.\ Rev.\ Lett.\  {\bf 86}, 1156 (2001)
[arXiv:hep-ex/0008065].

\bibitem{Sanders:2003ye}
M.~Sanders  [CDF Collaboration],
arXiv:hep-ex/0310033.

\bibitem{Traczyk:2002jh}
P.~Traczyk and G.~Wrochna,
arXiv:hep-ex/0207061; \\
C. Collard, M.-C. Lemaire, P. Traczyk, G. Wrochna,
CMS NOTE-2002/050

\bibitem{Martin:1998np}
A.~D.~Martin, R.~G.~Roberts, W.~J.~Stirling and R.~S.~Thorne,
Phys.\ Lett.\ B {\bf 443}, 301 (1998)
[arXiv:hep-ph/9808371]; 
arXiv:hep-ph/0307262.

\end{thebibliography}
\end{document}